\documentclass[12pt]{iopart}

\usepackage{iopams}
\expandafter\let\csname equation*\endcsname\relax
\expandafter\let\csname endequation*\endcsname\relax
\usepackage{amsmath, amsthm, amssymb}
\usepackage[overload]{textcase}
\usepackage{multirow}
\usepackage{rotating}
\usepackage{epsfig}
\usepackage{subfigure}
\usepackage{relsize}
\usepackage{times}
\usepackage{graphicx}

\newcommand{\screw}{$\frac{1}{2}\langle111\rangle$\@}
\newcommand\T{\rule{0pt}{2.6ex}}

\begin{document}

\title[Cereceda \etal]{Assessment of interatomic potentials for atomistic analysis of static and dynamic properties of screw dislocations in W}

\author{D.~Cereceda and J.~M.~Perlado}
\address{Universidad Polit\'ecnica de Madrid\\
28005 Madrid, Spain}
\author{A.~Stukowski, S.~Queyreau, and J.~Marian$^{\ast}$}
\address{Lawrence Livermore National Laboratory\\
Livermore, CA 94551}
\author{Lisa Ventelon and M.-C.~Marinica}
\address{Service de Recherches de M\'etallurgique Physique, CEA-DEN, F-91191, Gif-sur-Yvette, France}
\author{M.~R.~Gilbert}
\address{EURATOM/CCFE Fusion Association, Culham Science Centre, Abingdon, OX14 3DB, United Kingdom}

\eads{\mailto{$^{\ast}$marian1@llnl.gov}}

\begin{abstract}
  Screw dislocations in bcc metals display non-planar cores at zero temperature
  which result in high lattice friction and thermally activated strain rate behavior.
  In bcc W, electronic structure molecular statics calculations reveal a compact,
  non-degenerate core with an associated Peierls stress between 1.7 and 2.8 GPa. However, a full
  picture of the dynamic behavior of dislocations can only be gained by using more
  efficient atomistic simulations based on semiempirical interatomic
  potentials. In this paper we assess the suitability of five different potentials in terms of
  static properties relevant to screw dislocations in pure W. As well, we
  perform molecular dynamics simulations of stress-assisted
  glide using all five potentials to study the dynamic behavior of screw dislocations
  under shear stress. Dislocations are seen to display
  thermally-activated motion in most of the applied stress range, with a gradual
  transition to a viscous damping regime at high stresses.
  We find that one potential predicts a core transformation
  from compact to dissociated at finite temperature that affects the 
  energetics of kink-pair production and impacts the mechanism of motion.
  We conclude that a modified embedded-atom potential achieves the best
  compromise in terms of static and dynamic screw dislocation properties, although
  at an expense of about ten-fold compared to central potentials.
\end{abstract}

\maketitle

\section{Introduction.\label{sec:intro}}
Tungsten (W) is being considered as a leading candidate for plasma-facing
applications in magnetic fusion energy (MFE) devices. The most attractive
properties of W for MFE are its high melting point and thermal
conductivity, low sputtering yield and low long-term disposal radioactive
footprint. These advantages are accompanied unfortunately with very low
fracture toughness characterized by brittle trans- and inter-granular failure,
which severely restrict the useful operating temperature window \cite{zinkle2011}.

Transgranular plasticity in refractory metals, including W, is governed by the
temperature dependence of screw dislocation motion.
W is typically alloyed with 5$\sim$26 at.\% Re to increase low temperature ductility
and improve high temperature strength and plasticity \cite{Wbook}.  The physical
origins behind the Re-induced ductilization have been discussed in the
literature \cite{romaner2010,gludovatz2011,li2012} and point in some way or another to
alterations in the core structure of \screw~screw dislocations, which both reduce the
effective Peierls stress $\sigma_P$ and extend the number of possible slip pathways.
A direct consequence of a reduced Peierls stress, {\it e.g.}\ as via Re alloying,
is an enhanced dislocation mobility at low temperatures. Recent electronic
structure calculations of $\sigma_P$ in pure W give values between 1.7 and 2.8 GPa
\cite{romaner2010,samolyuk2013}. This means that, under most conditions relevant to technological
applications, where stresses are of the order of only a few hundred MPa, a reduction in
$\sigma_P$ of a few hundred MPa may not be significant to the plastic behavior of W and
W alloys. Instead, it is the thermally-activated and three-dimensional character
of screw dislocation motion, the associated solution softening behavior, as well as
the temperature dependence of the core structure, that control bulk ductility.

All of these aspects cannot be studied in atomistic detail using current experimental
capabilities. By contrast, atomistic methods based on semiempirical potentials have
enabled large-scale molecular dynamics (MD) simulations, so that, at present, calculations
of single-dislocation mobility, core structure and transformations, etc., can be
obtained with reasonable accuracy. However, care must be exercised when choosing
from the dozen or so W potentials available in the literature.
Semiempirical force fields with both pair and cohesive contributions
({\it e.g.} following the embedded atom method formalism) are typically considered to achieve an optimum
balance between efficiency and accuracy. These are typically
fitted to reproduce some basic bulk and defect properties such as lattice parameter,
elastic constants, vacancy formation energy, surface energies, etc., but generally not dislocation
properties. Of these, it is known that the screw dislocation core structure at 0 K
should be non-degenerate (also known as \emph{compact}), as revealed by density functional theory (DFT)
calculations \cite{romaner2010,li2012,samolyuk2013}.

Previous atomistic calculations on screw dislocations in W have been performed by
Mrovec \etal\ \cite{mrovec2007}, Fikar \etal\ \cite{fikar2009}
and Tian and Woo \cite{tian2004}. Mrovec \etal\ studied the dislocation core structure
and calculated the Peierls stress at 0 K using a tight-binding-based bond-order
potential (TB-BOP)\footnote{BOP potentials include non-central atomic interactions to represent
the effect of $d$-electrons in transition metals}.
They predicted a non-degenerate core structure
and a Peierls stress of 4.3 GPa. For their part, Fikar \etal\
studied core structures and energies of screw dislocations using
three different interatomic potentials, all of which display dissociated cores.
Lastly, Tian and Woo examined the mobility of screw
dislocations also with an embedded-atom potential that predicts a dissociated core
structure. They were able to obtain dislocation velocities at stresses above the
Peierls stress at 0 K. However, no systematic study of dislocation motion in W
at finite temperature has been conducted. Characterizing dislocation motion in the
stress-temperature space is important to parameterize the so-called mobility functions
used in higher-level methods such as dislocation dynamics. The purpose of the mobility functions is to provide
a quantitative measure of the response of dislocations to applied and internal stresses.

Unfortunately, one of the most important difficulties associated with such studies
is the scale-dependent nature of MD simulations, which require exceedingly high
strain rates to drive the system over time scales accessible computationally, of the
order of a a few tens of ns. Because of these limitations, MD is incapable at present of
properly capturing the thermally-activated motion of screw dislocations at low stresses.
However, MD simulations can still provide valuable input in intermediate-to-high stress
conditions and in situations where the deformation rates are high.
The objective of this paper is to compare five different interatomic potentials --that have not been
fitted against screw dislocation data-- and assess their performance in terms of static and dynamic screw
dislocation properties. By static properties we mean several reference parameters
at 0 K as obtained with DFT calculations. The dynamic behavior is evaluated in terms of
screw dislocation mobility as a function of stress and temperature. Due to the absence of
`reference' mobility data against which to compare the potentials, we will simply
draw several general conclusions based on the inter-comparison among potentials.

The paper is organized as follows. First, we discuss the distinctive features
of each potential and calculate the structure
of a screw dislocation core. The Peierls potential and the $\gamma$ surface are then calculated
and verified against existing DFT and TB-BOP calculations. Subsequently, we introduce the
computational setup for the dynamic mobility simulations and calculate dislocation velocities
as a function of temperature and stress. Subsequently, a study of the core trajectories
in the plane defined by the glide and normal directions is carried out.
We finish by analyzing the causes of the temperature-dependent behavior of each
potential and emphasizing the insufficiency of static calculations
to fully characterize dislocation motion at finite temperatures.

\section{\label{sec:comp}Computational details}

\subsection{\label{subsec:pot}Interatomic potentials}
Our calculations have been performed with the parallel MD code \textsc{LAMMPS}~\cite{lammps}.
Table \ref{tab:pot} gives basic information about the five different potentials considered here,
among which there are three embedded-atom
method (EAM) potentials, one Tersoff-Brenner-type bond-order potential (TF-BOP),
and one modified EAM (MEAM). Note that the TB-BOP used by Mrovec \etal\ \cite{mrovec2007}
was deemed not suitable for dynamics simulations by its authors \cite{mrovec-priv} and has
thus not been considered here.
Hereafter they are referred to in the text by the identifiers
given in the table header. This selection of W potentials, from the dozen or so
available in the literature, is not meant to be an implicit assessment of the quality
of those not employed here.
\begin{table}[ht]
\caption{Properties of potentials used: lattice parameter $a_0$, shear modulus $\mu$, Peierls stress $\sigma_P$, computational cost, core structure at 0 K, and thermal expansion coefficient $\alpha$. Potentials EAM1, EAM2 and TF-BOP display a threefold symmetric (degenerate) core, while EAM3 and MEAM predict compact (non-degenerate) cores. The values of the volumetric thermal expansion coefficients, $\alpha$, are used in Section \ref{mark}. 
} 
\begin{minipage}{15cm}
\centering 
\begin{tabular}{c c c c c c} 
\hline\hline 
Potential \T & EAM1 & EAM2 & EAM3 & TF-BOP & MEAM \\ [0.5ex] 
\hline 
Ref.\T & \cite{zhou2001} & \cite{ackland} & \cite{cea} & \cite{bop} & \cite{meam}\\ [1ex] 
$a_0$ (\AA) & 3.165 & 3.165 & 3.143 & 3.165 & 3.188 \\
$\mu$ (GPa)\footnote[2]{W is isotropic elastic and, thus, the value of $\mu$ given is equally valid for $\{110\}$ and/or
$\{112\}$ slip.} & 160 & 163 & 161 & 170 & 161 \\
$\sigma_P$ (GPa)\footnote[3]{For consistency, our Peierls stress calculations use the same geometry as the DFT calculations by Romaner \etal~\cite{romaner2010} and Samolyuk \etal~\cite{samolyuk2013}, which reveal a value of $\sigma_P$ between 1.7 and 2.8 GPa.} & 4.0 & 2.0 & 1.8 & 1.1 & 3.2 \\ [1ex]
\parbox{2.5cm}{Computational cost relative to EAM1} \T & 1.0 & 0.4 & 0.9 & 5.4 & 9.1 \\
& \multirow{3}{*}{\includegraphics[width=2cm]{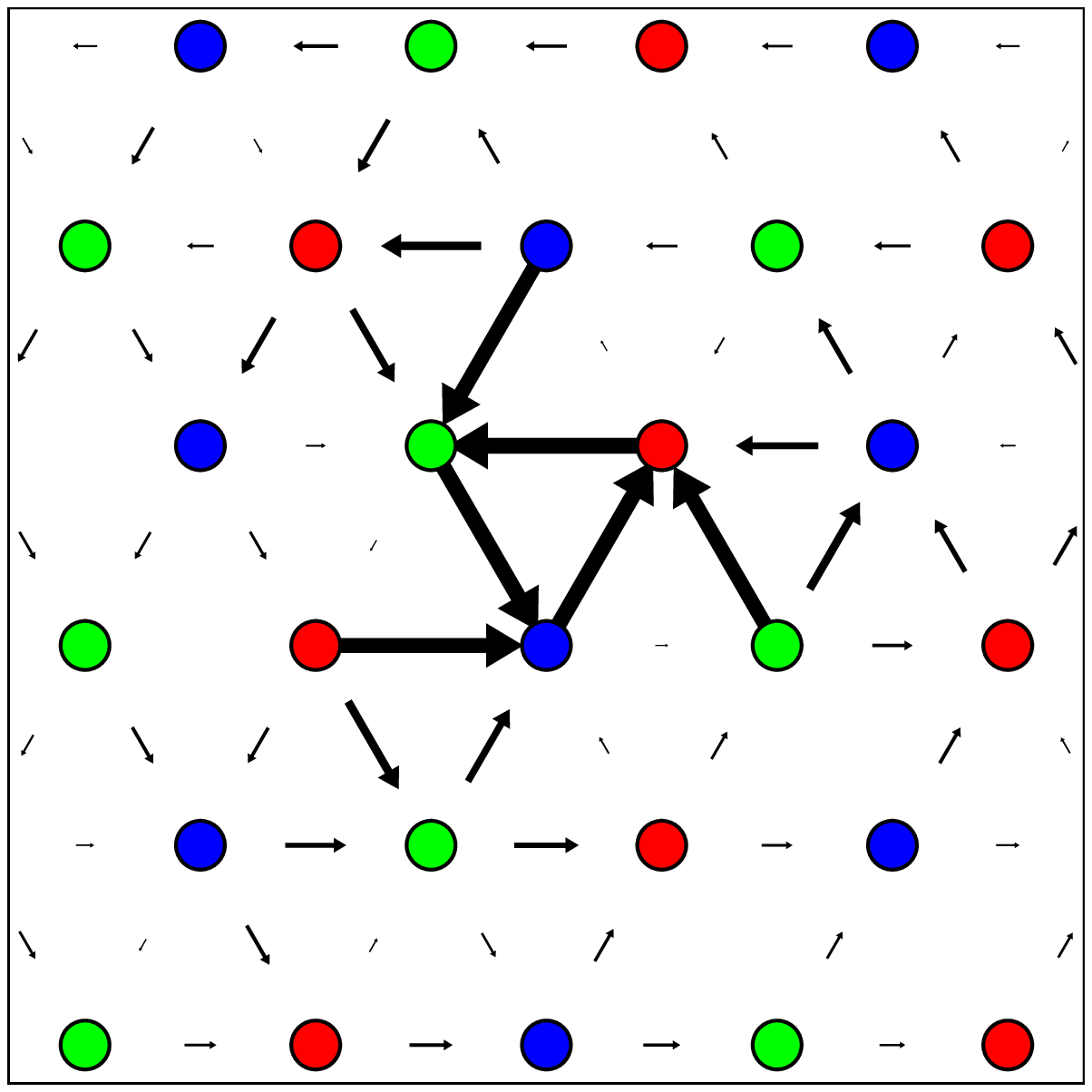}} & \multirow{3}{*}{\includegraphics[width=2cm]{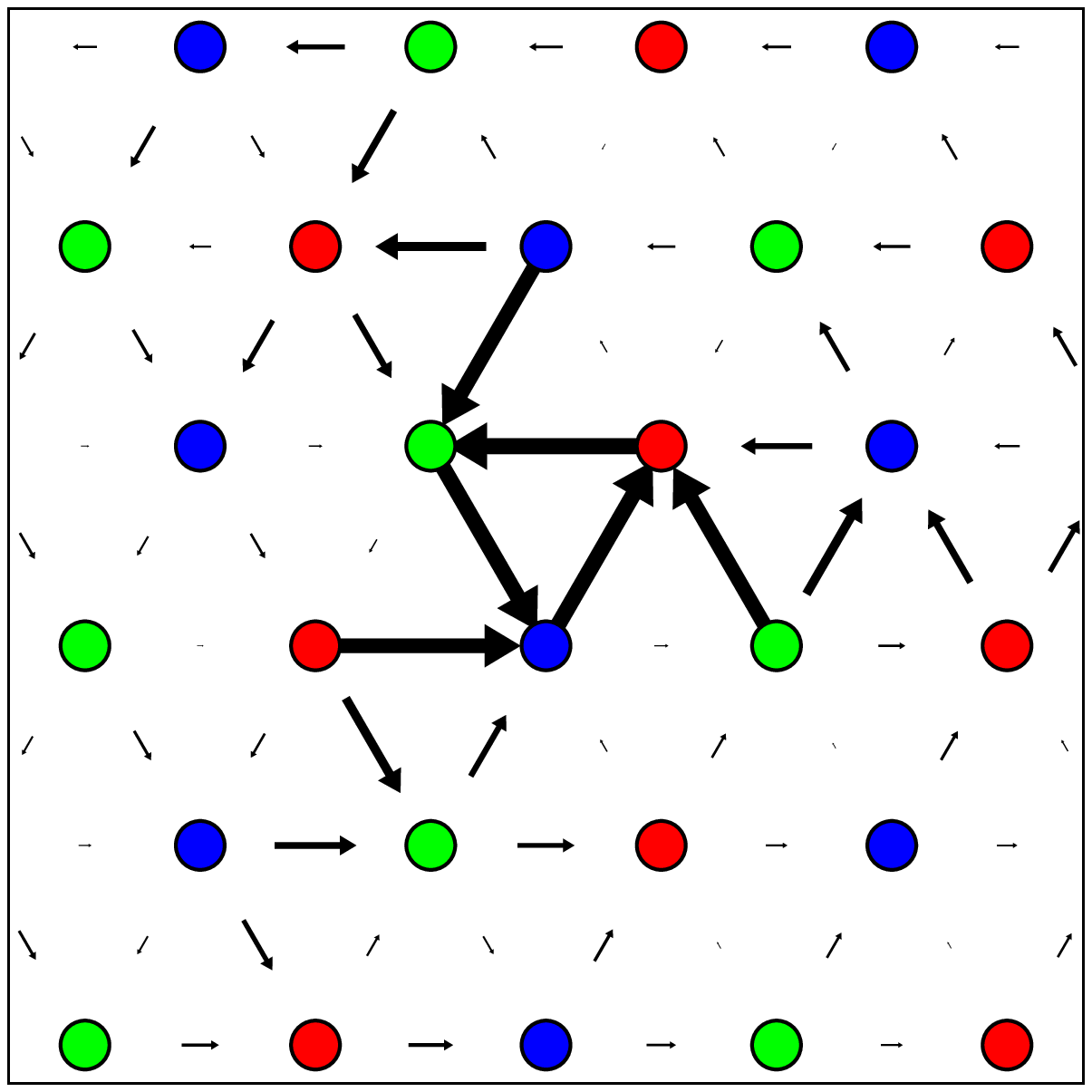}} & \multirow{3}{*}{\includegraphics[width=2cm]{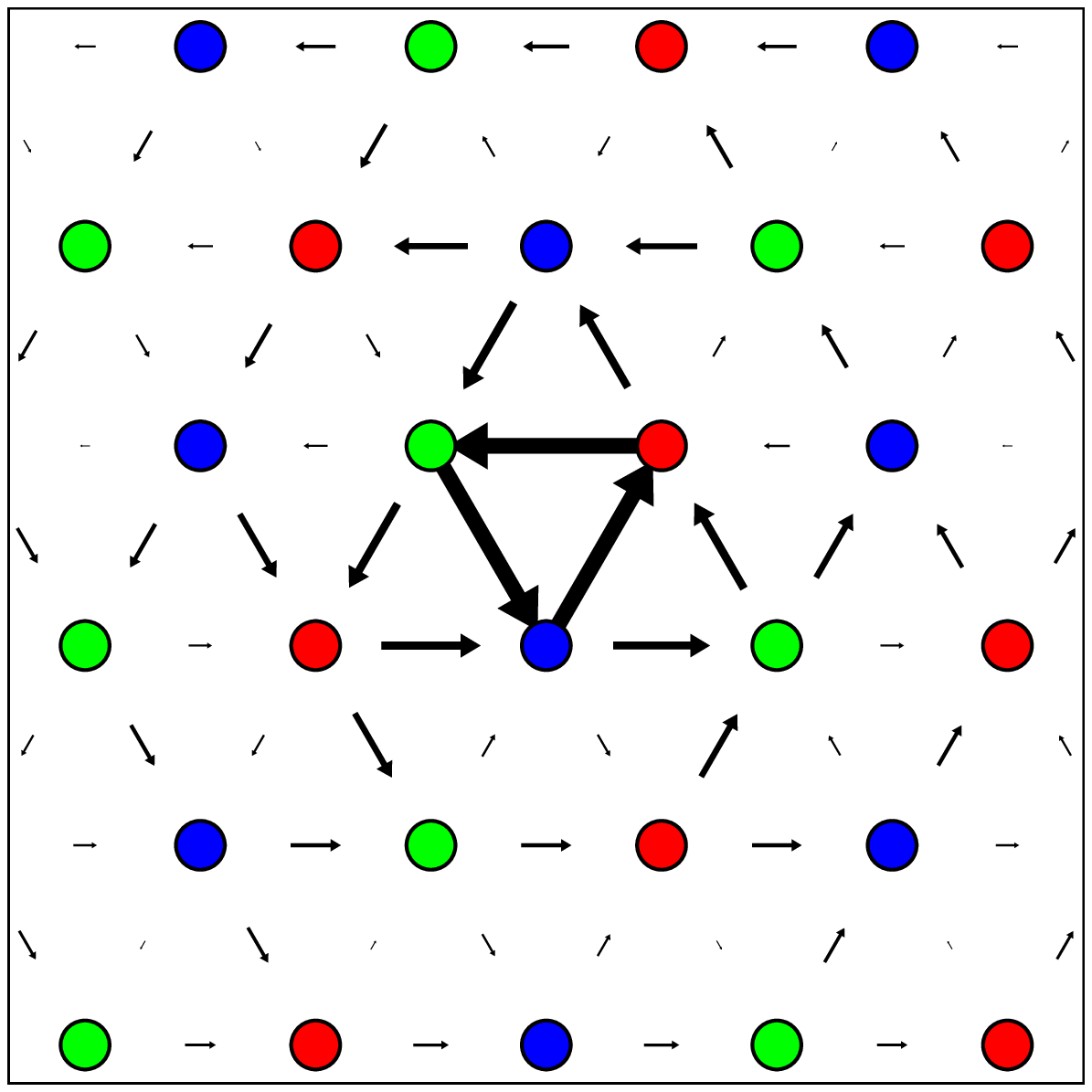}} & \multirow{3}{*}{\includegraphics[width=2cm]{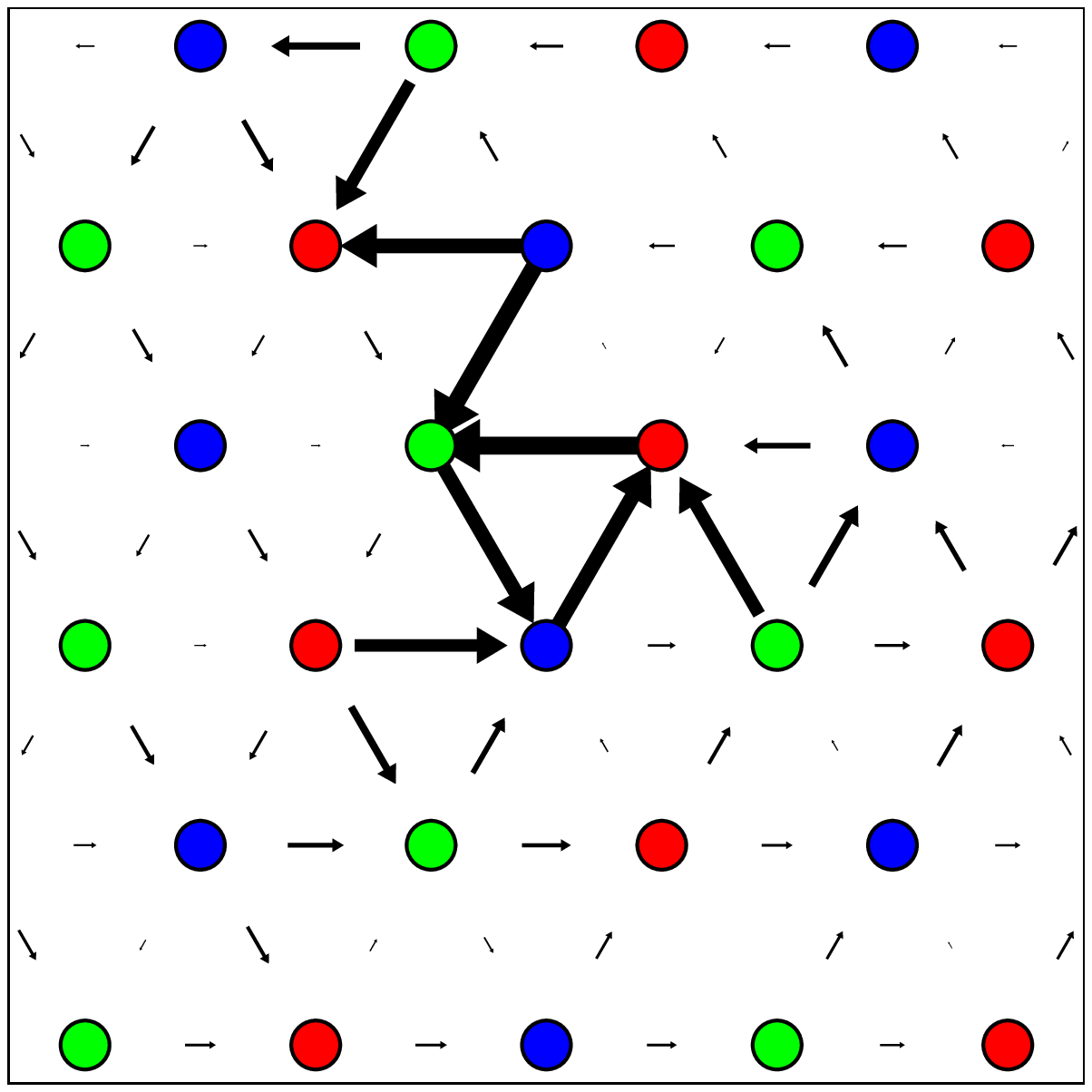}} & \multirow{3}{*}{\includegraphics[width=2cm]{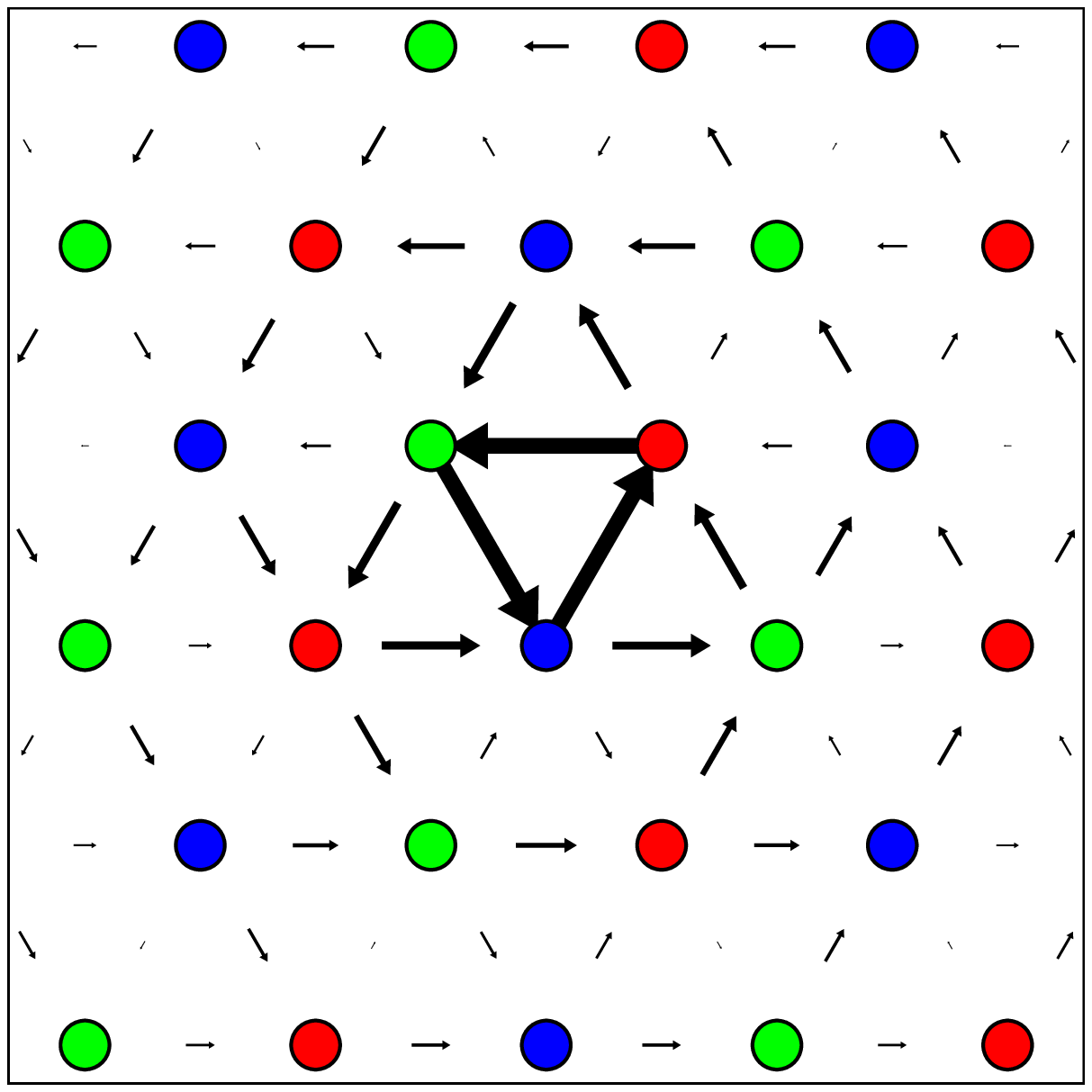}} \\
\parbox{2.5cm}{Core structure at 0 K\footnote[4]{DFT calculations predict a compact, non-degenerate core.}} & & & & & \\
& & & & & \\
& & & & & \\
$\alpha$ ($\times10^{-5}$ K$^{-1}$)\footnote[5]{The experimental value for $\alpha$ is 1.45 to $1.91\times10^{-5}$ K$^{-1}$ in the 1000 to 2000 K temperature interval \cite{knibbs1969}.} & 1.40 & 2.42 & 1.76 & 2.38 & 1.64 \\
\hline 
\end{tabular}
\renewcommand{\footnoterule}{}
\end{minipage}
\label{tab:pot} 
\end{table}

\begin{figure}[h]
  \centering
  \includegraphics[width=12cm]{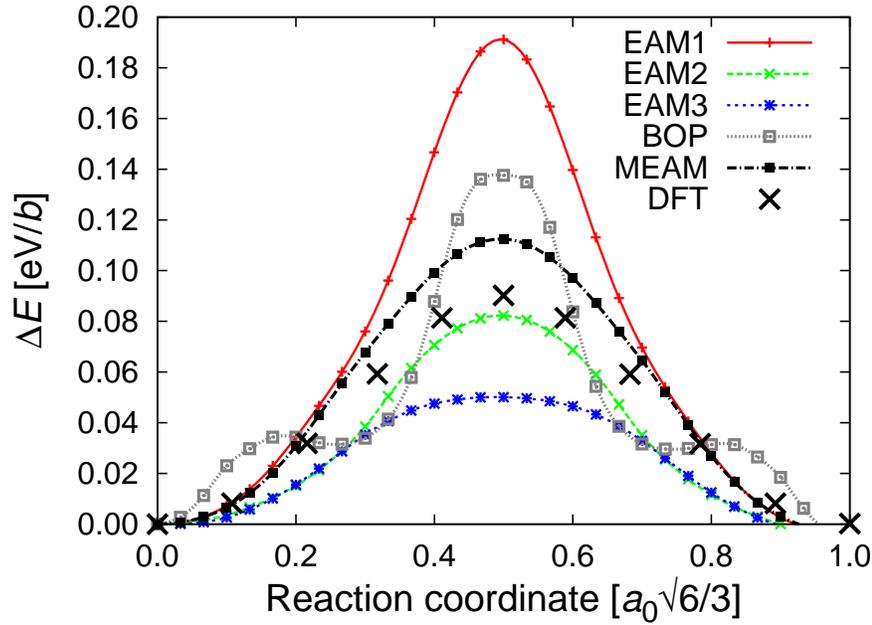}
  \caption{Peierls potential for all potentials tested here.
    DFT calculations from Ventelon \etal~\cite{ventelon2012} are shown for comparison.}
  \label{peierls}
\end{figure}

\begin{figure}[h]
  \centering
  \includegraphics[width=12cm]{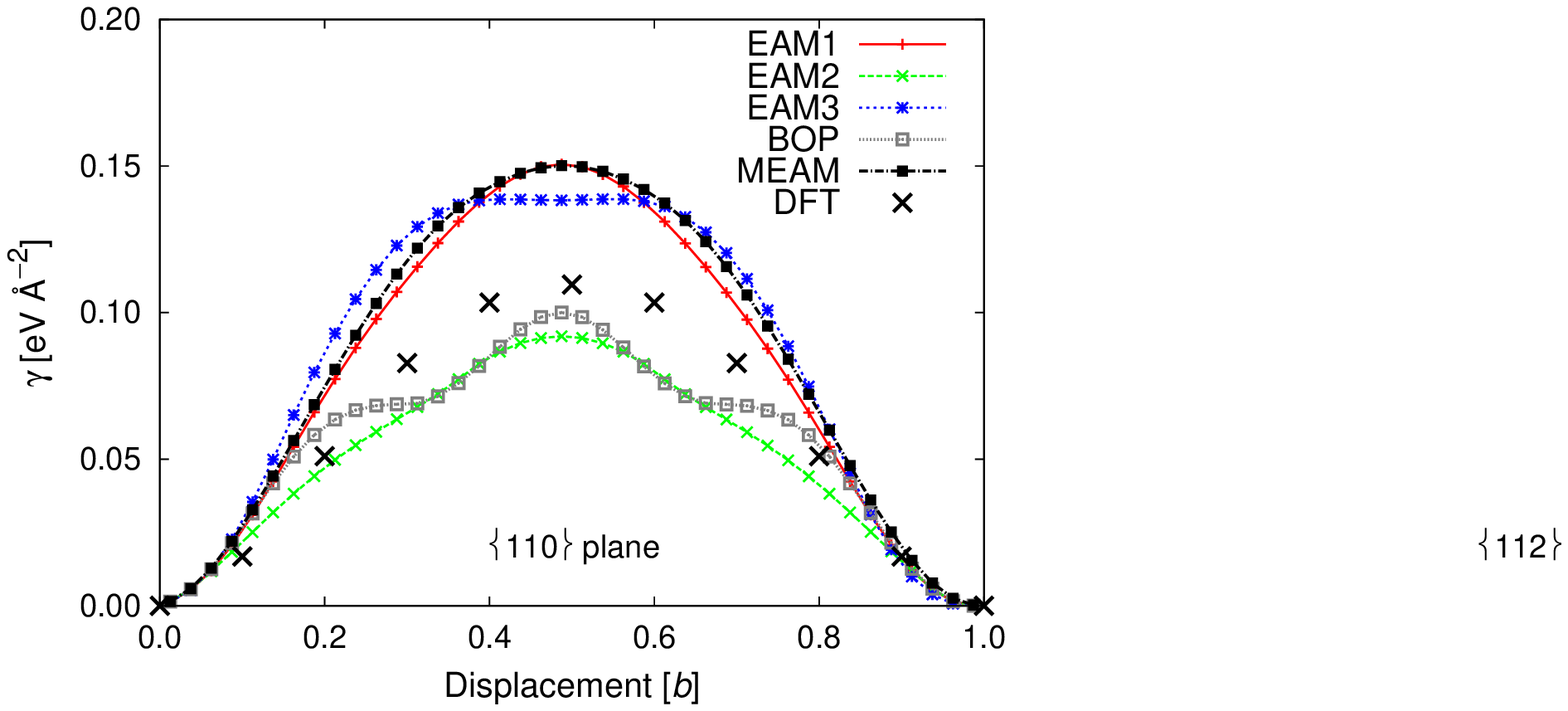}
  \caption{$(1\bar{1}0)$ $\gamma$ surface along the $[111]$
    direction for all the potentials considered in this work.
    DFT calculations from Ventelon \etal~\cite{ventelon2012} are shown for comparison.}
  \label{fig:gamma}
\end{figure}

Two important quantities for characterizing screw dislocation cores at 0 K are the
Peierls potential, defined as the energy path from one equilibrium position
to another on a $\{110\}$ plane, and the
$\gamma$ surface along the $[111]$ direction also on $\{110\}$ planes.
The Peierls potential governs the morphology of kinks (e.g.~\cite{gordon2010}) while workers such as Duesbery and Vitek
\cite{duesbery1998} have provided evidence for a direct correspondence between the shape of the
\screw$\{110\}$ gamma surface and the screw dislocation core structure.
These are plotted, respectively, for all potentials in Figs.\ \ref{peierls} and
\ref{fig:gamma} on the $(1\bar{1}0)$ plane. DFT data for both calculations are also shown for comparison.

The Peierls potential was obtained using the nudged elastic band method
\cite{neb} in the manner proposed by Gr\"oger and Vitek \cite{arxiv-groger},
whereas the DFT calculations in both cases were obtained using a \emph{plane wave
self-consistent field} code as described in Refs.\ \cite{ventelon2007} and \cite{ventelon2012}.

\subsection{\label{subsec:cell}Simulation setup}
To measure dislocation velocities, we have performed stress-controlled simulations
of \screw~dislocations with the maximum resolved shear stress (MRSS) on a $\{112\}$ plane.
The justification to focus on $\{112\}$-type planes is twofold. First, as Argon and Maloof
\cite{argon1966} have shown, under tensile loading, most loading orientations
and temperatures, result in some degree of \screw$\{112\}$ slip.
Also, Li \etal~\cite{li2012} have shown that $\{112\}$ slip is important
in W alloys with high Re concentrations.
Second, certain EAM potentials intrinsically deviate from MRSS behavior
when the MRSS plane is of the $\{110\}$ type for reasons that have
been discussed at length in the literature \cite{chaussidon2006,groger,gilbert2011}.

We have provided the details regarding the computational setup for this type
of simulations in prior publications \cite{gilbert2011,cereceda2012}.
Suffice it to say that orthogonal boxes of sufficient size with axes
$x\equiv\frac{1}{2}[111]$, $y\equiv[1\bar{1}0]$, and $z\equiv[11\bar{2}]$,
corresponding to the line, glide, and plane normal directions,
respectively, were used to mitigate finite size effects (cf.\ Ref.~\cite{gilbert2011}).
$\sigma_{xz}$ is then applied on the computational cell boundaries and
simulations are conducted in the $NPT$ ensemble.
Periodic boundary conditions are used in the line and glide directions.
The reference cell dimensions were $L_x=25\left[\frac{\sqrt{3}}{2}a_0\right]$,
$L_y=100\left[\sqrt{2}a_0\right]$, and $L_z=50\left[\sqrt{6}a_0\right]$, where the amounts in brackets are the
dimensions of the nominal unit cell in the coordinate system employed here.
The reference configuration contains $7.5\times10^5$ atoms, which
results in strain rates of $1.4\times10^{6\sim7}$ s$^{-1}$ for dislocation
velocities between 10 and 100 m$\cdot$s$^{-1}$. 

All simulations were run on LLNL's \textsc{ATLAS} cluster using 128 and 256 processors
at a reference cost of $\approx10^{-7}$ CPU seconds per atom per time step for potential EAM1.

\section{\label{sec:results}Results}
The simulation setup, boundary conditions, and velocity calculations, as they relate to the present
work, are discussed in depth by Cereceda \etal~\cite{cereceda2012}.
The temperature and stress ranges covered were, respectively, 300 to 2100 K and 200
to 2000 MPa. All the simulations were run for 100 ps and configuration data were
extracted every picosecond.
The procedure to extract dislocation velocities from MD simulations
is well established in the literature \cite{bul&cai,olmsted2005,gilbert2011}:
from the position of the core, velocities are
calculated as the derivative of the displacement-time curves for each case.

\subsection{\label{subsec:mob}Screw dislocation mobility}
\begin{figure}[h]
  \centering
  \subfigure{
    \label{eam1}
    \includegraphics[width=7cm]{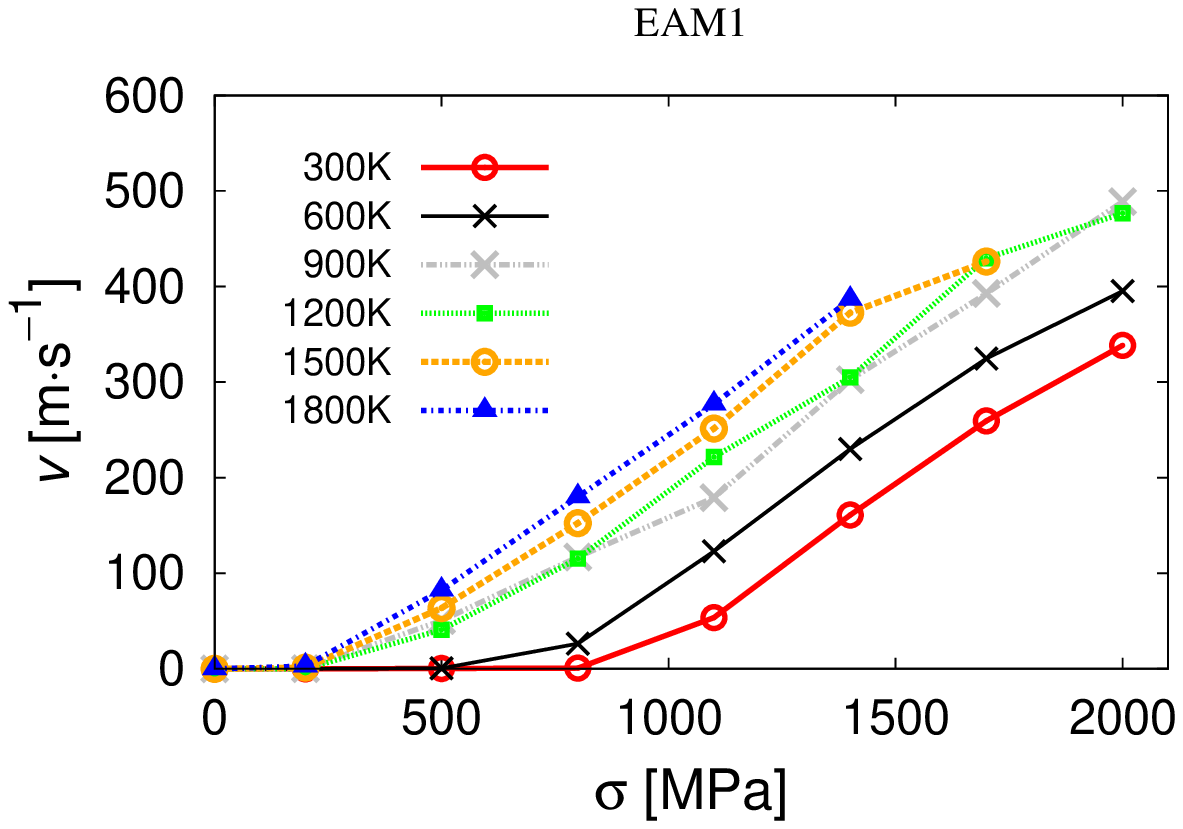}
  }
  \subfigure{
    \label{eam2}
    \includegraphics[width=7cm]{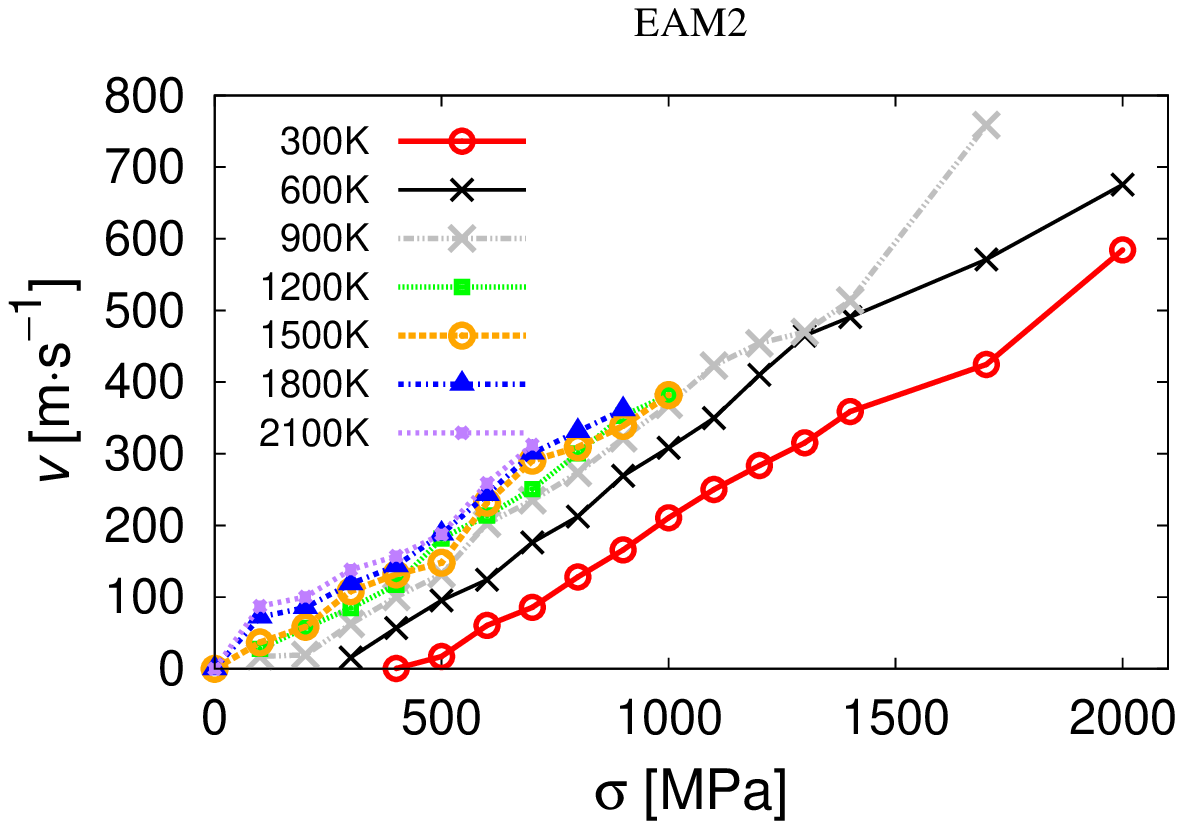}
  }
  \subfigure{
    \label{eam3}
    \includegraphics[width=7cm]{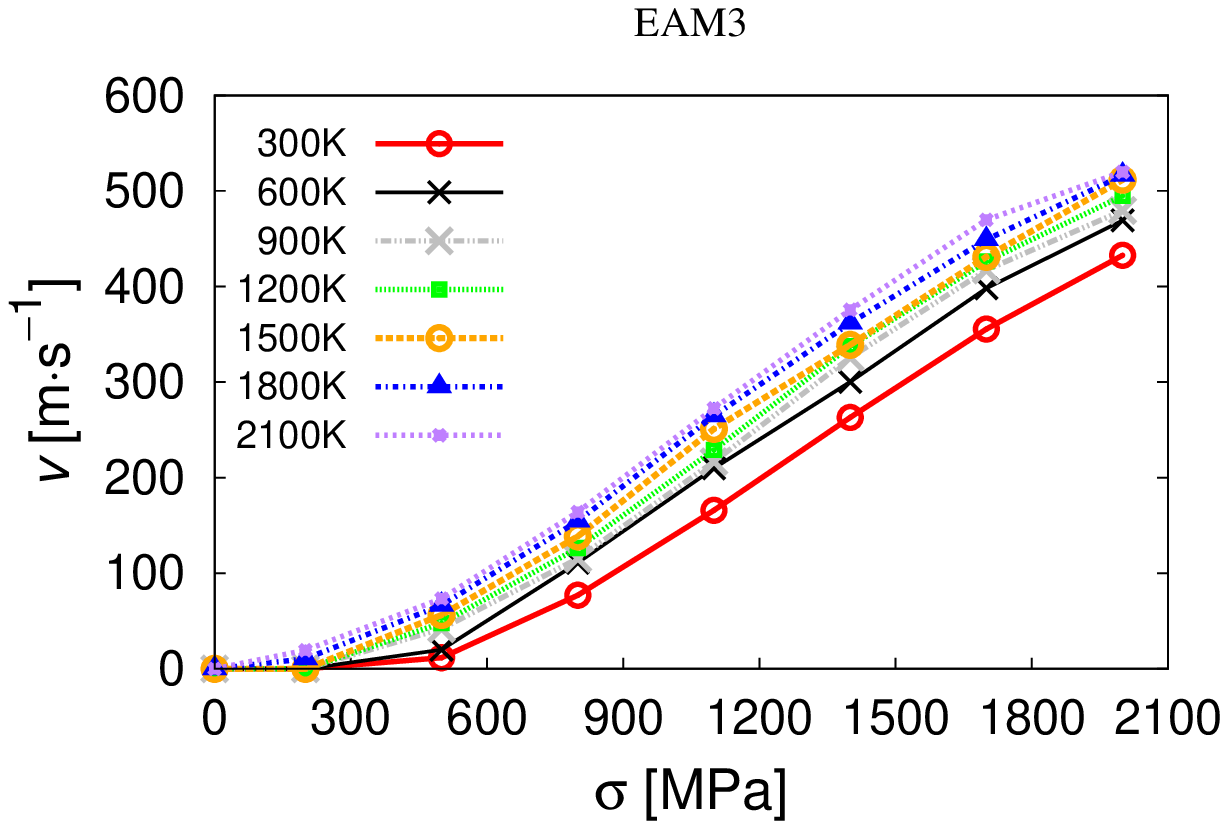}
  }
  \subfigure{
    \label{bop}
    \includegraphics[width=7cm]{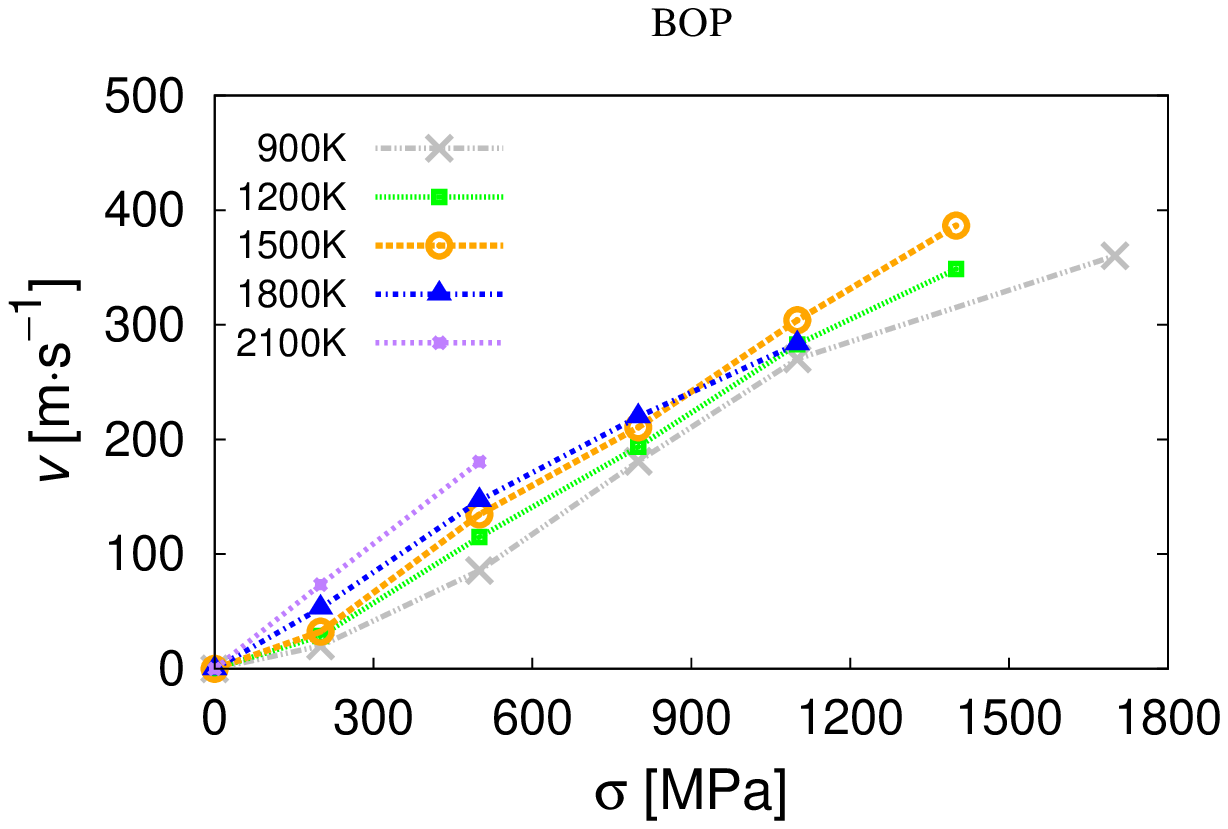}
  }
  \subfigure{
    \label{meam}
    \includegraphics[width=7cm]{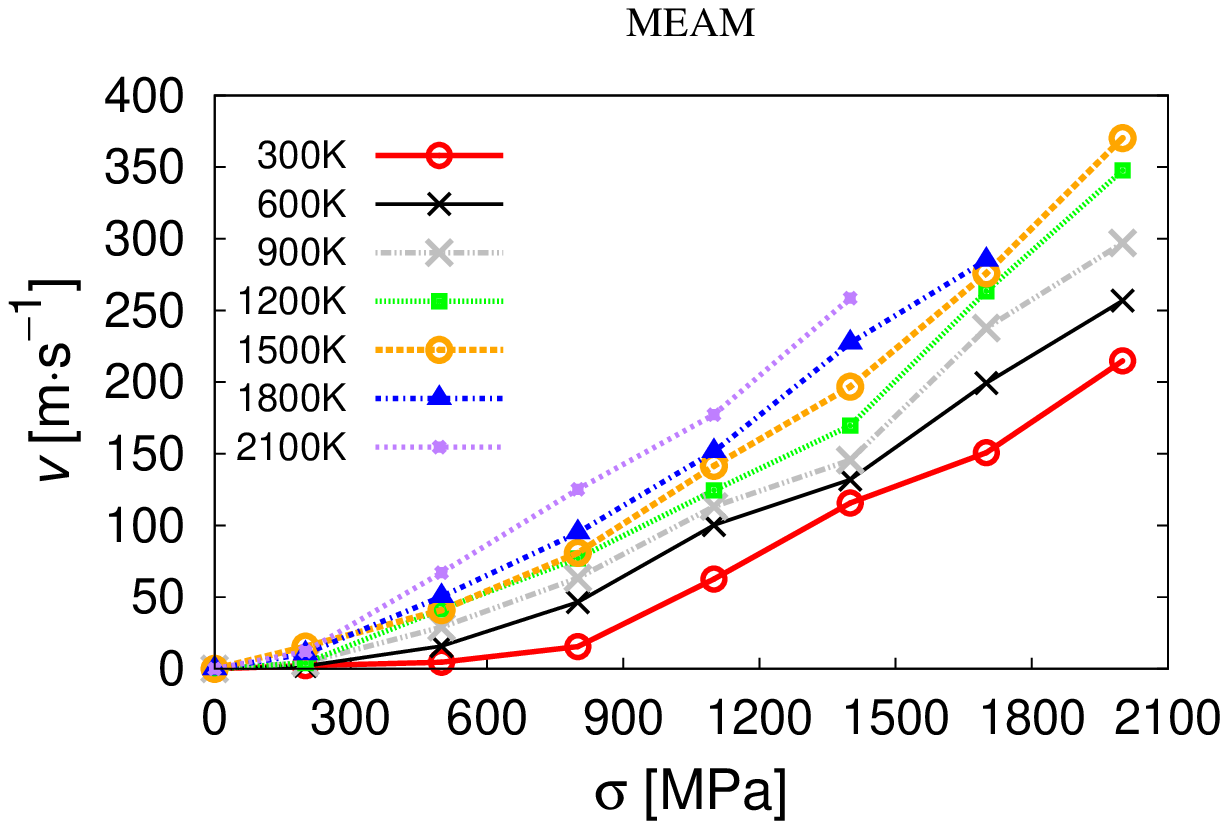}
  }
  \caption{Screw dislocation velocity as a function of applied shear stress
    and temperature for the five interatomic potentials considered here. Note
    that the velocity and stress axes are not on the same scale for each case.}
  \label{fig:moball}
\end{figure}

Figure \ref{fig:moball} shows all the ($\sigma$-$v$) data for the five
interatomic potentials tested. The figures also contain the temperature
dependence for each case.
Generally, the velocities increase monotonically with stress and temperature, although
at different rates depending on the potential.
To first order, the mechanism of motion followed by the dislocations depends on the
Peierls stress. This means that, at a maximum applied shear stress of 2000 MPa, the
EAM1, EAM2, and MEAM potentials both operate under $\sigma_P$ (cf.\ Table \ref{tab:pot}),
while for the EAM3 and TF-BOP
there are several data points above it. In either case, dislocation motion is mostly
governed by the thermally activated kink-pair nucleation mechanism, and thus display
an exponential dependence with $\sigma$ and $T$. This can be qualitatively
appreciated in the figure, although in the Appendix a more quantitative analysis is carried out.

\begin{figure}[h]
  \centering
  \subfigure{
    \label{traj600}
    \includegraphics[width=10cm]{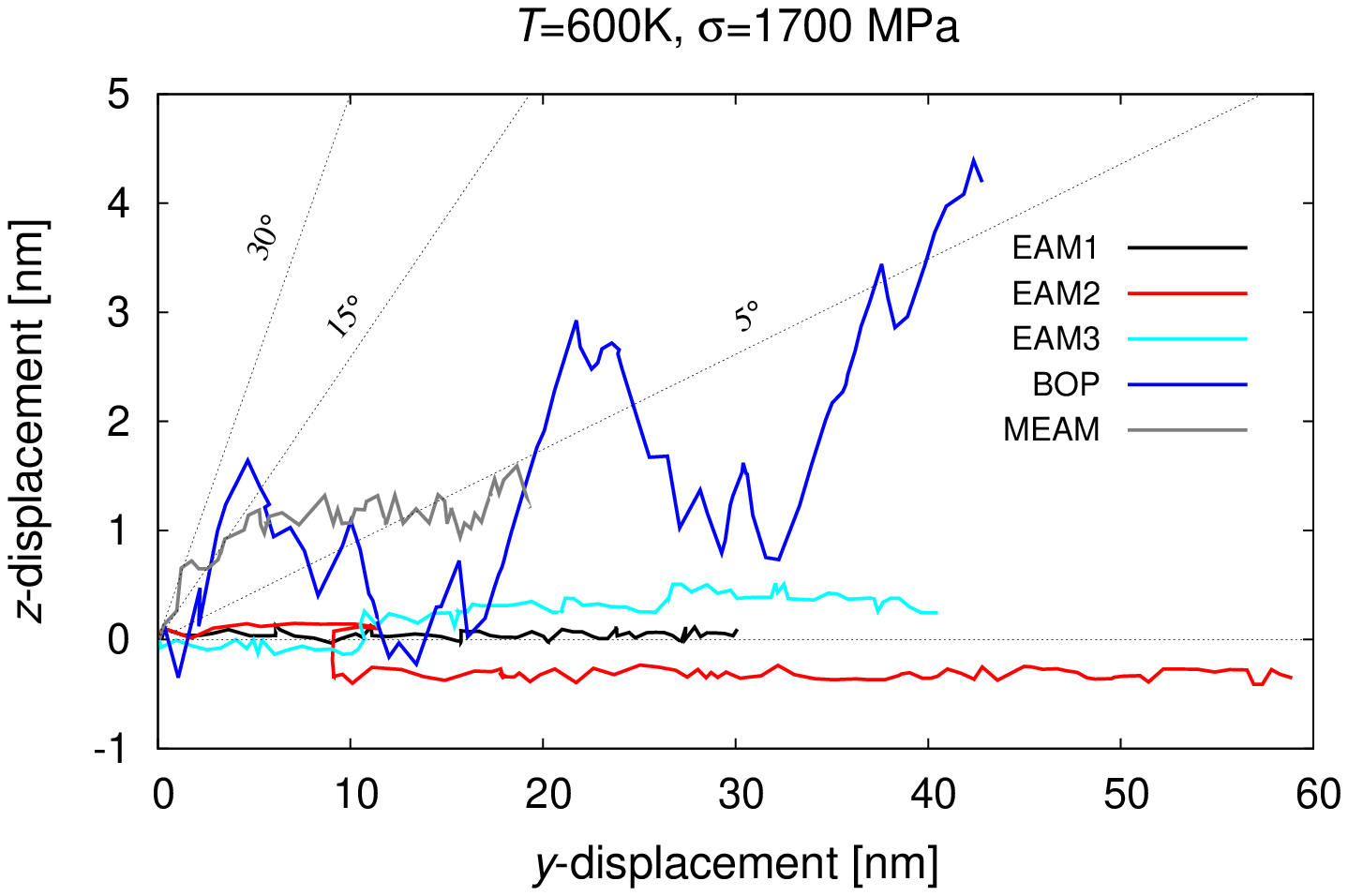}
  }
  \subfigure{
    \label{traj1200}
    \includegraphics[width=10cm]{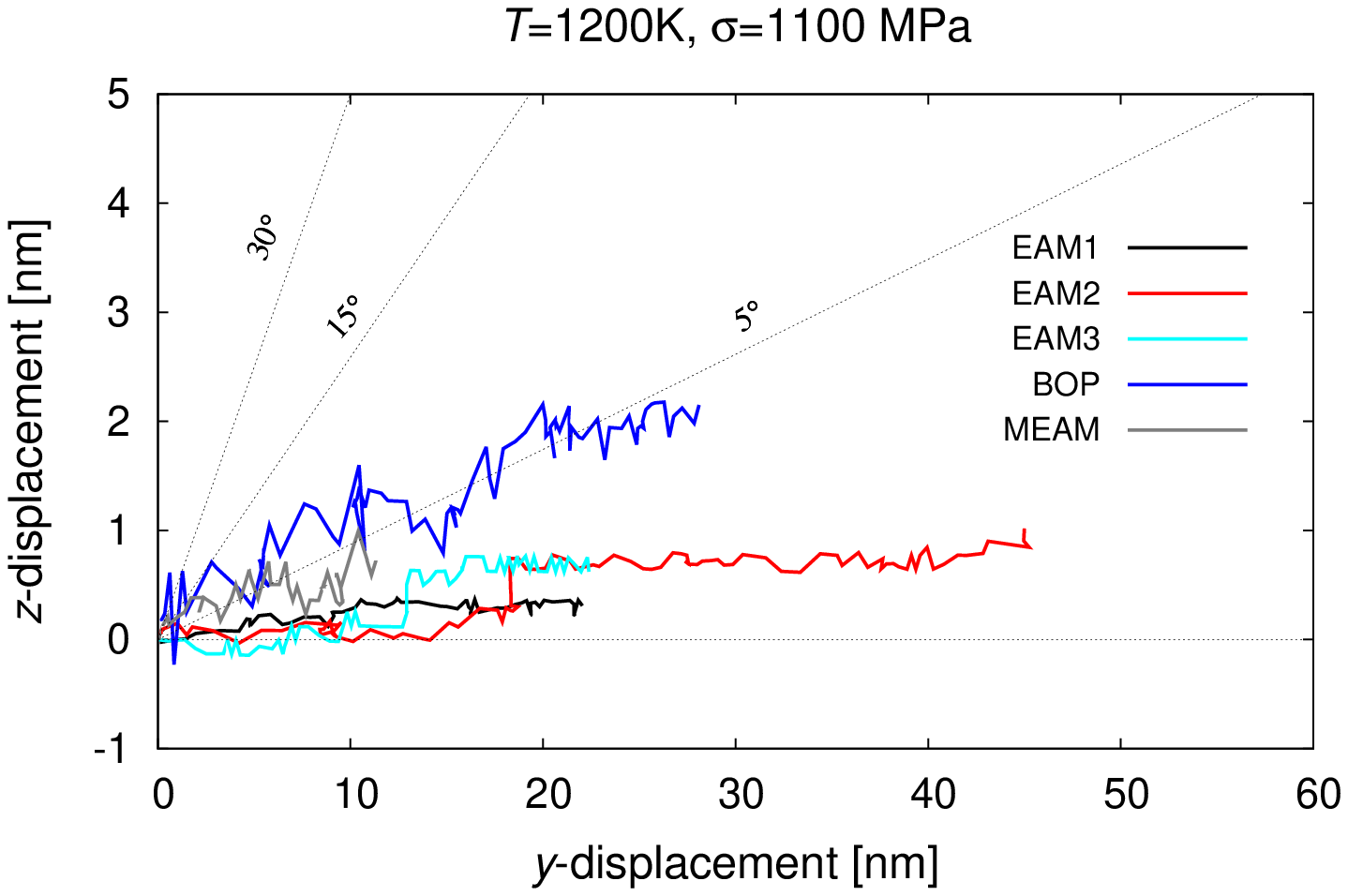}
  }
  \caption{Line-averaged dislocation trajectories on the $yz$-plane for two combinations
    of $\sigma$ and $T$ and over 100 ps. Planes forming 0, 5, 15 and $30^{\circ}$ with the
    $(11\bar{2})$ MRSS plane are represented with dotted lines (angles not to scale).
    Except for the TF-BOP potential, all the simulations yield small $<$5$^{\circ}$ deviations
    from MRSS motion.}
  \label{fig:traj}
\end{figure}

Another important aspect of dislocation motion is the extent of MRSS motion
displayed, {\it i.e.} whether there are deviations from glide on the MRSS $\{112\}$-type plane.
In Figure \ref{fig:traj} we analyze the trajectories on the $yz$-plane for
different combinations of $\sigma$ and $T$ over 100 ps of simulation.
Perfect MRSS behavior is characterized by trajectories parallel to $0^{\circ}$.
As the figure shows, all the EAM potentials display nearly perfect MRSS behavior,
while for the MEAM small deviations in the acceleration phase are captured.
The TF-BOP potential displays the most erratic motion with an overall deviation of the order
of five degrees. At higher temperatures and stresses, this effect is enhanced to the
point that the dislocation exits the simulation box only a few picoseconds after
the shear stress is applied. This is the reason why there are fewer data points in the
$\sigma$-$v$ curves shown in Fig.\ \ref{fig:moball} for the TF-BOP potential.
The trajectories shown in the figure are \emph{effective}, {\it i.e.} they
are not sufficiently time resolved to capture the atomistic details of
dislocation motion. Nevertheless, the operating mechanism of motion is by way of
nucleation and propagation of kink pairs on $\{110\}$ planes adjacent to the MRSS
$(11\bar{2})$ plane.

\subsection{\label{subsec:core}Dislocation core structure at finite temperature}
\begin{figure}[h]
  \centering
  \subfigure{
    \label{300K}
    \includegraphics[width=7cm]{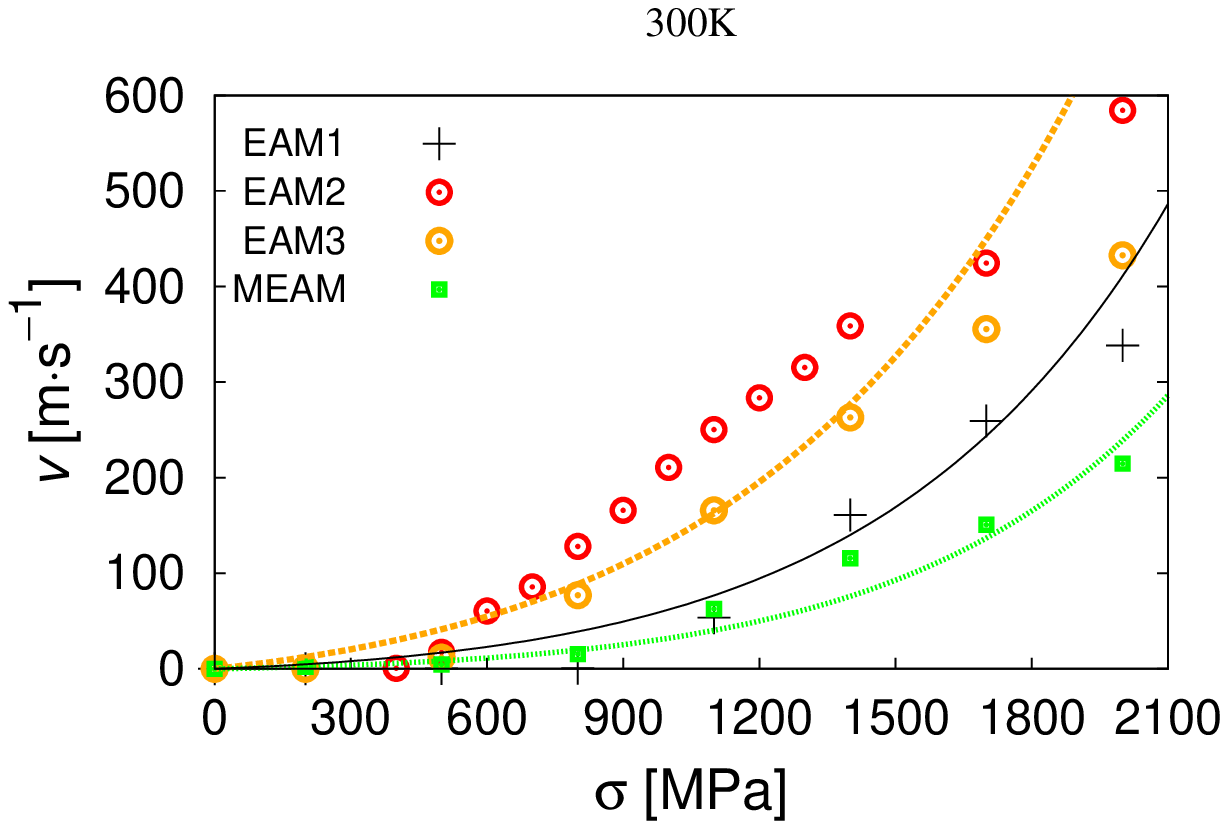}
  }
  \subfigure{
    \label{600K}
    \includegraphics[width=7cm]{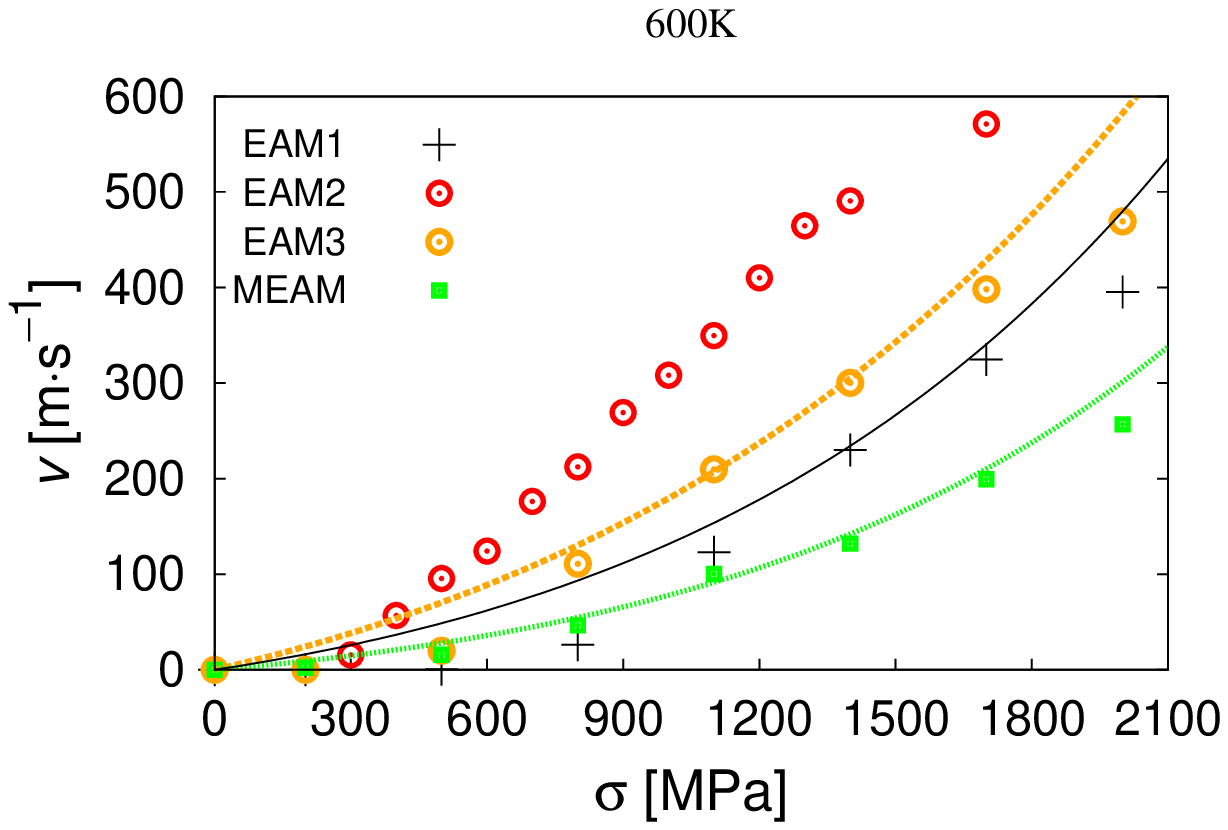}
  }
  \subfigure{
    \label{900K}
    \includegraphics[width=7cm]{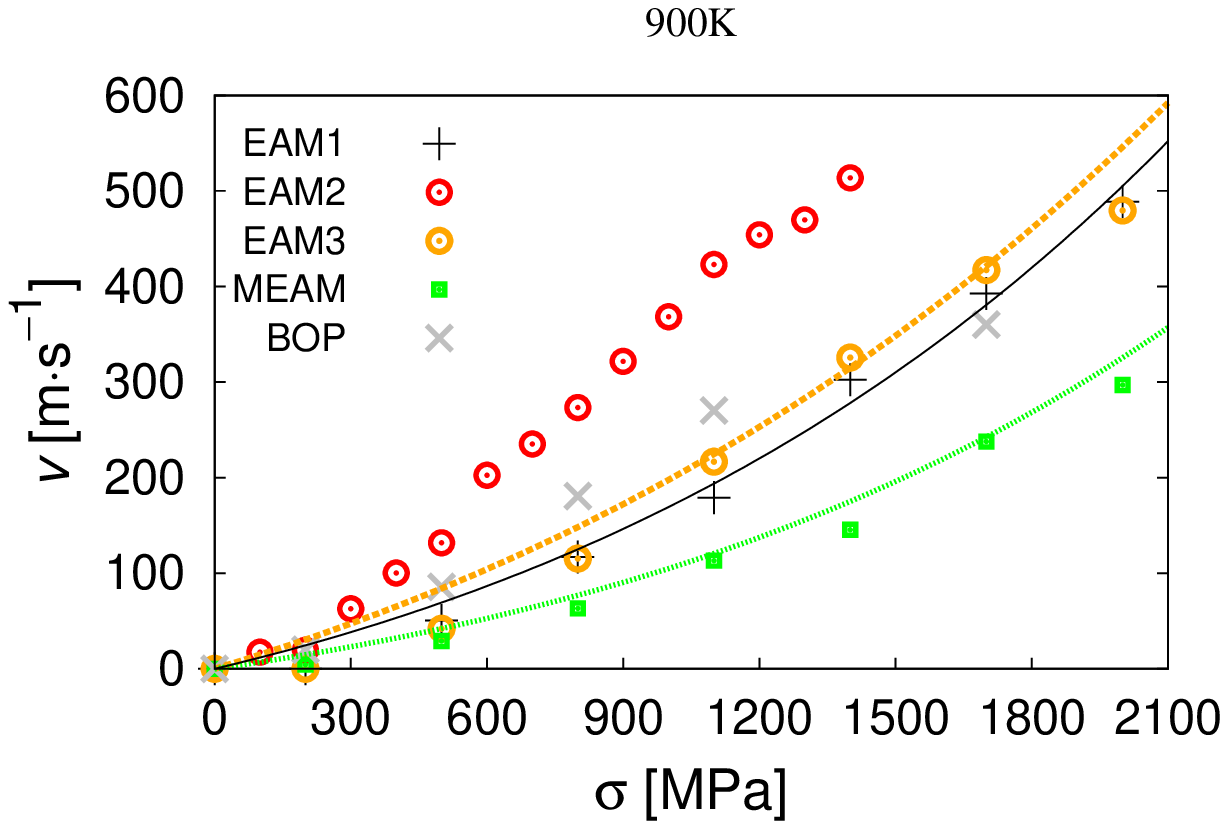}
  }
  \subfigure{
    \label{1800K}
    \includegraphics[width=7cm]{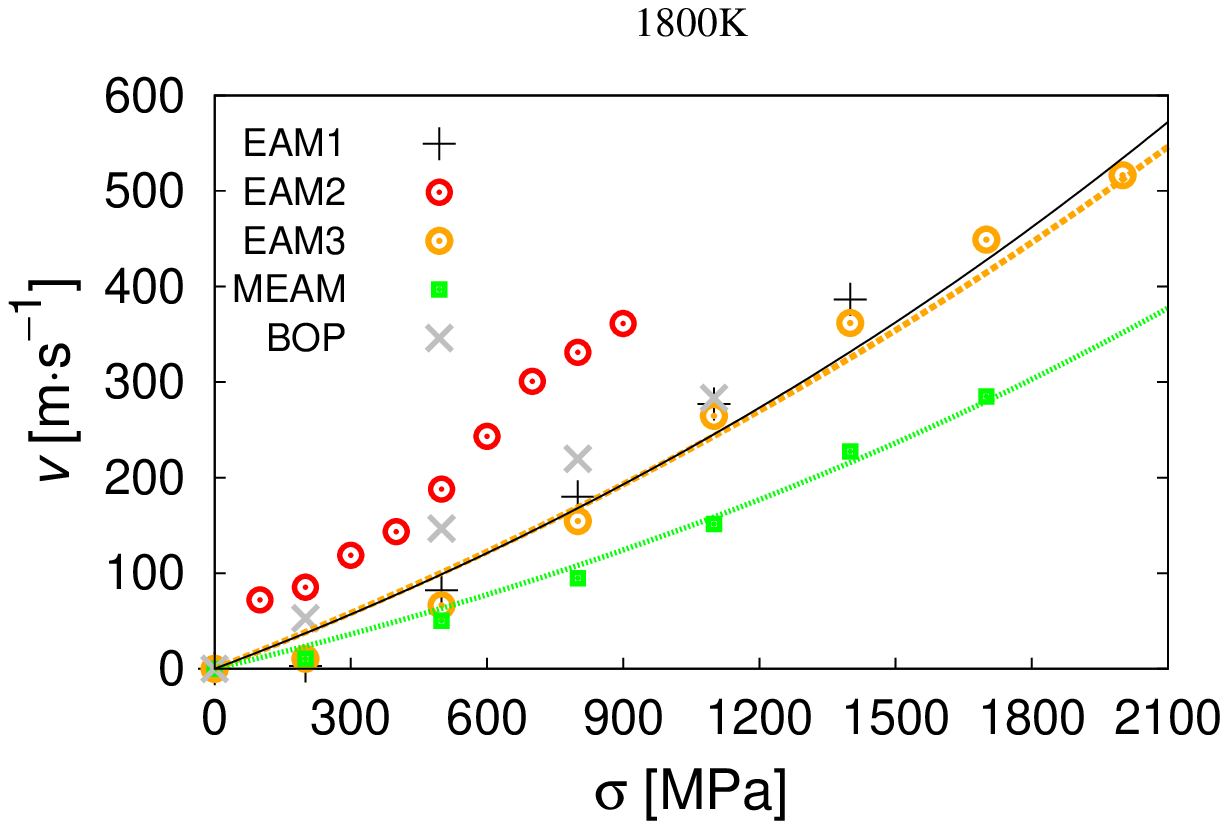}
  }
  \caption{Comparison of interatomic potentials for the data given in Fig.\ \ref{fig:moball}.
    The colored lines correspond to exponential fits obtained in the Appendix for
    potentials EAM1, EAM3 and MEAM.}
  \label{fig:allpot}
\end{figure}

\begin{figure}[h]
  \centering
  \includegraphics[width=14cm]{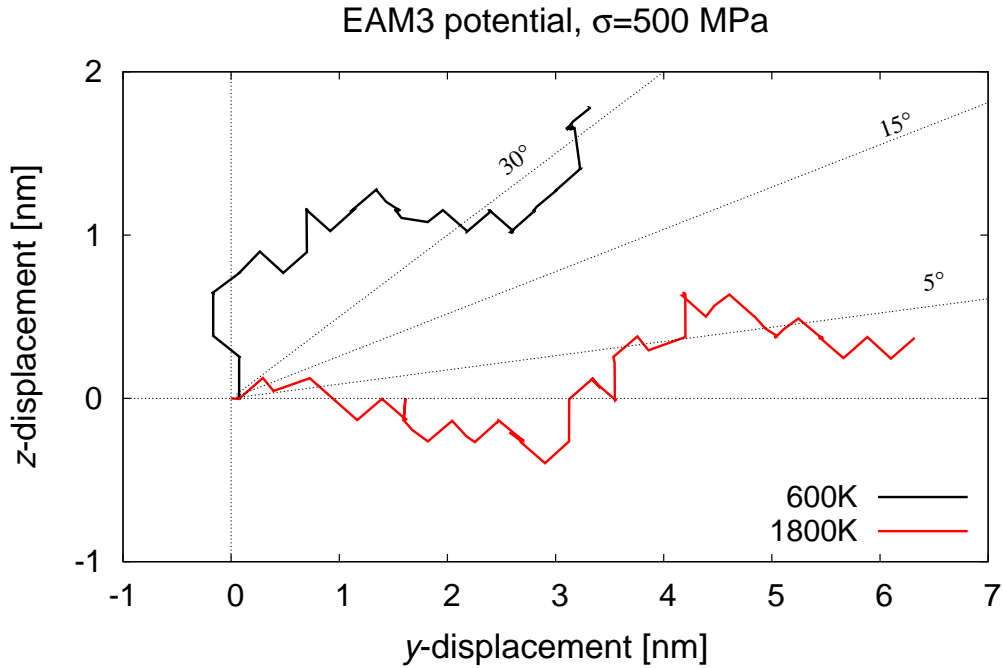}
  \caption{Line-averaged dislocation trajectories on the $yz$-plane for the EAM3
    potential at 500 MPa. Results for two 200-ps temperatures above and below the
    presumed core transformation
    temperature of around 1200 K are shown. Planes forming 0, 5, 15 and $30^{\circ}$ with the
    $(11\bar{2})$ MRSS plane are represented with dotted lines (angles not to scale).}
  \label{fig:random}
\end{figure}

As shown in Fig.\ \ref{fig:moball}, the $\sigma$-$v$ data are not conducive to
comparison among potentials. Instead, in Fig.\ \ref{fig:allpot} they are plotted as a function
of interatomic potential for a number of selected temperatures. The figure reveals an
interesting trend: the \emph{relative} behavior of all the potentials remains unchanged
for all temperatures with the exception of EAM3. At low temperatures,
this potential exhibits a relatively high dislocation mobility, akin to that
displayed by `fast' potentials such as EAM2. However, above 900 K, the mobility is
reduced (relative to the other interatomic potentials)
to values more in line with `slower' potentials such as EAM1.
Moreover, if one examines the trajectories followed by the dislocation at 500 MPa
\footnote{To be meaningful, this analysis must be performed at relatively low stresses
to interfere the least amount possible with the investigated temperature effect.},
a notable difference in behavior within the EAM3 potential can be observed.
At a temperature of 600 K, the dislocation follows a biased path on an effective
glide plane forming $\approx$30$^{\circ}$ with the MRSS plane. However, at 1800 K,
the dislocation follows a path that deviates only slightly from that dictated by
the Peach-K\"ohler force ({\it i.e} $0^{\circ}$).
This is quantitatively displayed in Figure \ref{fig:random}, where this time the
trajectories are resolved with atomistic detail.
The figure shows unequivocally that dislocation motion proceeds via the formation of
kink pairs on $\{110\}$ planes bordering the MRSS $[11\bar{2}]$ plane (at
$\pm30^{\circ}$). Moreover, the details of the trajectory at 600 K suggest biased
formation on the $(10\bar{1})$ plane ($+30^{\circ}$), whereas at 1800 K \emph{random-walk} behavior
is displayed, with kink pairs forming equally on both available $\{110\}$ planes.

The behaviors illustrated in Figs.\ \ref{fig:allpot} and \ref{fig:random} for potential EAM3 suggest a change in
core structure with temperature for
a given stress state \footnote{We know that stress also induces its own core transformations
as explained in Ref.\ \cite{rodney2008}.}.
To examine the physical structure of the dislocation core at different temperatures
one can use time-averaged differential displacement (DD) maps (these maps were used in
Table \ref{tab:pot} for each 0 K configurations).
\begin{figure}[h]
  \centering
  \includegraphics[width=12cm]{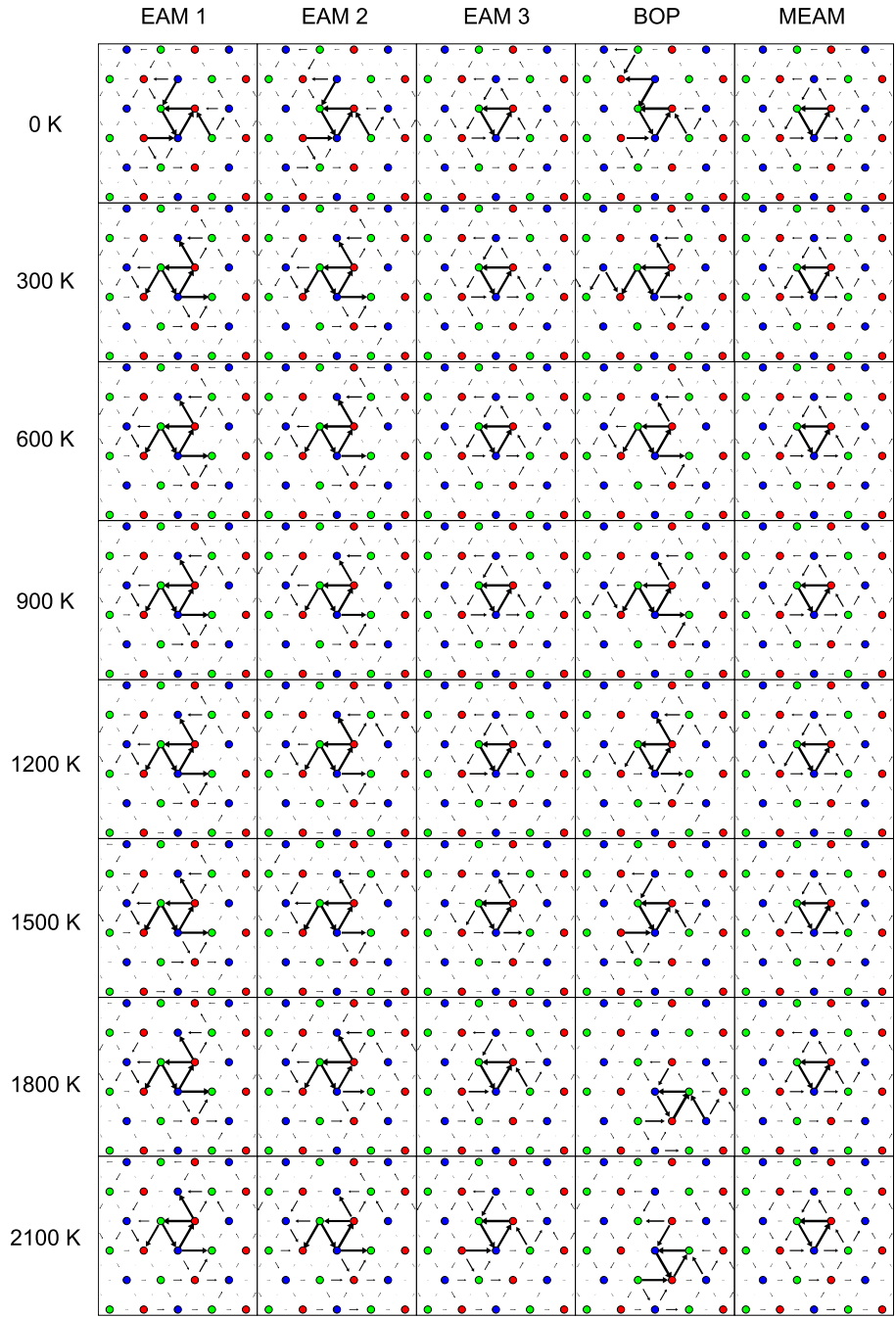}
  \caption{Time-averaged core structures for the five potentials tested here
  in the entire temperature range.}
  \label{fig:core}
\end{figure}
The DD maps are obtained by running MD simulations of
crystals containing four screw dislocations arranged in a balanced quadrupole
configuration and periodic boundary conditions. The size of the simulation box
is $20\times15\times18$ multiples of the bcc lattice vectors
$[111]\times[\bar{1}2\bar{1}]\times[\bar{1}01]$. The dimensions are adjusted
to the equilibrium lattice constant at the given temperature. For the finite
temperature simulations, the displacement of each atomic string is determined
by averaging over all 40 atoms in the string and over a time window of 100 fs, being
sufficiently long to avoid noise due to thermal vibrations yet short enough to not
capture diffusive behavior. The results are shown in Fig.\ \ref{fig:core}
for configurations in the $0<T<2100$ K interval. The figure confirms that the the EAM3
core is the only one showing an appreciable transformation from non-degenerate to degenerate,
clearly seen at and above 1500 K. Although DD maps are a useful tool to
quickly analyze core structures, next we complement the results in Fig.\ \ref{fig:core}
with a more quantitative approach based on fundamental lattice properties.

\subsection{Analysis of screw dislocation core stability.}\label{mark}

Duesbery and Vitek \cite{duesbery1998} have provided a simple rule that relates the
shape of the \screw$\{110\}$ $\gamma$-surface to the core structure at 0 K.
They used the following inequality:
$$\gamma\left(\frac{b}{3}\right)>2\gamma\left(\frac{b}{6}\right),$$
to predict whether a screw dislocation will display a compact core.
$\gamma\left(\frac{b}{3}\right)$ and $\gamma\left(\frac{b}{6}\right)$ are the
energies corresponding to the $\frac{b}{3}$ and $\frac{b}{6}$ magnitudes of the
generalized fault vectors, which can be obtained by reference to Fig.\ \ref{fig:gamma}.
The idea is that, if the above inequality is satisfied, $\frac{b}{6}$-type
faults will be preferred over $\frac{b}{3}$ ones, leading to non-dissociated core structures.
However, although Duesbery and Vitek applied this simple rule to six different
bcc metals\footnote{V, Cr, Nb, Mo, Ta, and W, all described by Finnis-Sinclair
potentials \cite{fs}} with remarkable success, we find that in our case it does not hold
for potentials EAM1 and EAM2. Thus, here we try a different approach based on the
analysis carried out by Gilbert and Dudarev \cite{gilbert2010}.

These authors have shown that, alternatively, the \screw~screw dislocation core structure in bcc systems can be related to the periodic interaction energy between adjacent \(\langle 111\rangle\) strings of atoms forming the crystal. Their analysis, which was performed primarily to help guide potential development, provides a framework to predict whether the favored core structure at 0 K is compact or dissociated. In particular, they derive the so-called ``first-nearest-neighbour (1NN) inter-string interaction law'' of the potential and use this in a 2D Frenkel-Kontorova (2D-FK) model of interacting \(\langle111\rangle\) strings to find the minimum energy screw core-structure. Here we extend their methodology to finite temperatures using the quasiharmonic approximation, {\it i.e.}~by relating volume changes to temperature via pre-computed thermal expansion coefficients for each potential. In this fashion, we first compute the inter-string interaction laws as a function of the lattice parameter and then obtain the equivalent temperature as: $T=3\left(a/a_0-1\right)/\alpha$. Here, $\alpha$ is the thermal expansion coefficient (given for each potential in Table \ref{tab:pot}), $a_0$ the lattice parameter at 0 K, and $a$ the lattice parameter corresponding to a temperature $T$ (within the quasiharmonic approximation). For the reminder of this section, we refer to $a$ as \(a(T)\) to highlight this temperature dependence.

\begin{figure}[h]
{\includegraphics[width=1.0\linewidth,clip=true,trim=0cm 0cm 1cm
1.8cm]{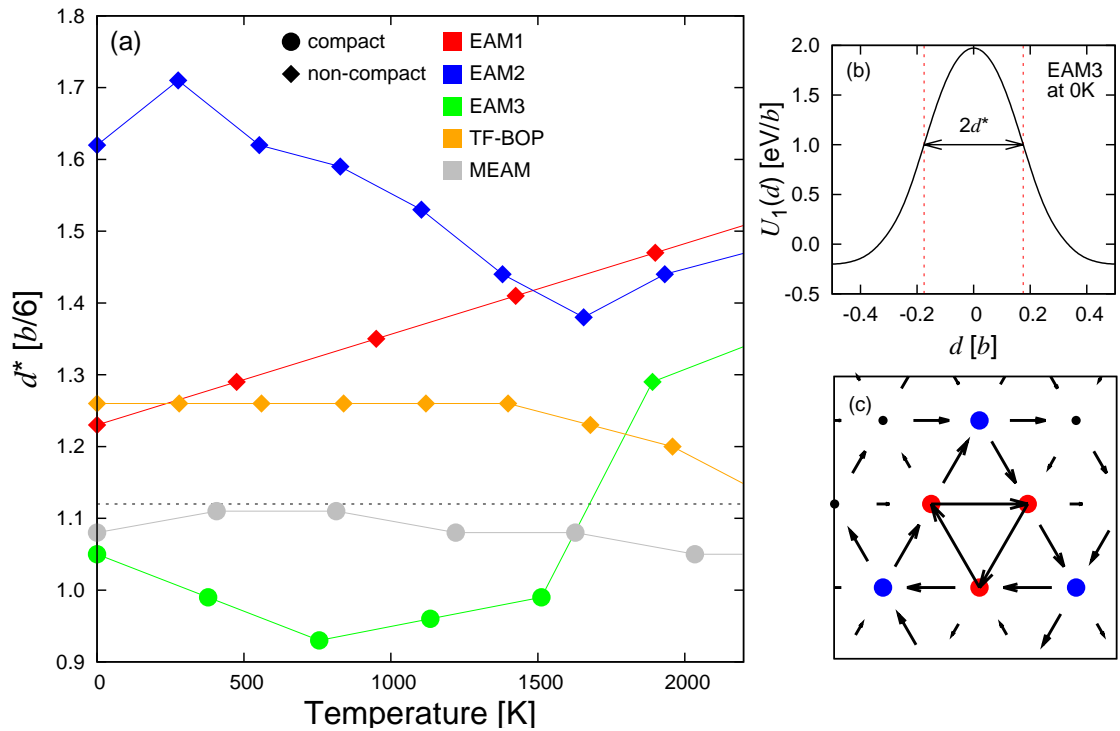}}
\caption{(a) Analysis of the favored core structure for the five different potentials as a function of temperature. A compact core is designated by a circle, and a non-compact by a diamond. The $y$-position of each point is the magnitude of distance of the inflection points in the 1NN string-interaction law ({\it i.e.} $d^{\ast}$, which is the distance at which $U^{\prime\prime}_1(d)=0$) normalized to the quantity \(b(T)/6\), where \(b(T)=\sqrt{3}a(T)/2\). The dashed horizontal line separates the region of phase space where compact cores are favored (below) from the region where non-compact are favored (above). (b) The 1NN string-interaction law for EAM3 at 0 K. The red dashed vertical lines indicate the position of the inflection points in this curve. (c) Differential displacement map of the compact core predicted at 0 K for the $U_1$ function from EAM3. Each of the three strings closest to the core (red circles) are separated from their two closest secondary strings (blue circles) by \(b/6\). In the figure the temperature dependence of $b$ is omitted for the sake of clarity.
}\label{core_atability_withT}
\end{figure}

For each temperature \(T\) the 1NN inter-string interaction law \(U_1(d)\) was derived by rigidly translating a single \(\langle 111\rangle\) string with respect to a perfect lattice with lattice parameter \(a(T)\) and measuring the associated variation in energy under the particular interatomic potential. The resulting curve, which, according to the 2D-FK model defined in Ref.\ \cite{gilbert2010}, is dominated by the contributions from the moving string interacting with its six 1NNs, can be unfolded using a Fourier analysis to produce the required pair-wise interaction law for the 2D-FK model. An example of such a law for EAM3 at 0 K is shown in figure~\ref{core_atability_withT}(b). A perfect screw dislocation, inserted into a lattice of \(\langle111\rangle\) atomic strings (for a given \(a\equiv a(T)\)) according to the isotropic elasticity solution, was then relaxed using \(U_1(d)\) and the nature of the relaxed core was determined by visual inspection of its differential displacement map.

Figure~\ref{core_atability_withT}(a) shows the variation in the favored core structure as a function of \(T\) for each of the five potentials. On the $y$-axis of the plot we have calculated the ratio of the string separation \(d^{\ast}\) associated with the inflection points in the corresponding \(U_1(d)\) law (highlighted by the vertical dotted lines in figure~\ref{core_atability_withT}(b)) to \(b(T)/6\), where \(b(T)\) is the corresponding Burgers vector of each potential as a function of \(T\).

As observed by Gilbert and Dudarev \cite{gilbert2010}, the favored core structure depends on the position of the inflection points of the $U_1(d)$ function. Specifically, a fully compact core is characterized by minimum string separations of \(b(T)/6\), which are the in-line separation distances between each of the three \(\langle111\rangle\) strings immediately surrounding the core (red circles in Figure~\ref{core_atability_withT}(c), which shows a differential displacement map for a compact core) and their two nearest strings forming the next shell of strings out from the core (blue circles in \ref{core_atability_withT}(c)). When the inflection points in \(U_1(d)\) are located at a distance of less than or equal to \(b(T)/6\), then the compact, non-dissociated core is always stable.
Furthermore, even if the separations \(d^{\ast}\) associated with \(U^{\prime\prime}_1(d)=0\) are such that \(|d^{\ast}|\) is somewhat greater than \(b(T)/6\), the compact core may still be stable provided that \(U_1\) is only slowly varying around these inflection points ---meaning that the forces (\(-U_1^\prime (d)\)) between strings are relatively constant over a range of \(d\) values. This, for example, is the situation in the case of the MEAM potential in Fig.~\ref{core_atability_withT}(a), where the ratio $d^{\ast}/(b(T)/6)$ is greater than one for all \(T\), but the favored core remains compact.

However, when the ratio is significantly greater than one, as is the case for both EAM1 and EAM2 at all temperatures, then the compact core becomes unstable and the secondary strings out from the core tend to move towards (along $\langle111\rangle$) one of their primary-string neighbors (signified by the major arrows in the ``arms'' of the non-compact core shown for EAM1 in Table \ref{tab:pot}) ---ultimately leading to the stabilization of the non-compact, three-fold symmetric dissociated core.

Thus, if, as a function of \(T\), there is significant variation in this ratio, then the preferred equilibrium core structure can also change. In our analysis, we find that, consistent with the transition observed in Fig.\ \ref{fig:core}, there is a large shift for EAM3 in the value of $d^{\ast}/(b(T)/6)$ ($d^{\ast}$ is such that $U^{\prime\prime}_1(d^{\ast})=0$) above \(\sim1500\) K. At this point, the equilibrium core structure diverges from the compact core, and becomes more and more dissociated as temperature increases further.
In the next section we discuss the implications of our findings.

\section{\label{sec:disc}Discussion}

\subsection{Comparison of static properties of potentials}
The five interatomic potentials tested here follow different formulations
and have been fitted to different physical properties. It is not our objective
to discuss the fitting process or the quality of each one, but only to discuss
their performance in relation to screw dislocation modeling.
When selecting potentials for screw dislocation simulations,
two of the properties most looked at are the core structure and the Peierls stress.
For W, these have been obtained using electronic structure calculations of different sorts,
which reveal a compact core and $\sigma_P$ between 1.7 and 2.8 GPa.
Regarding the core structure at 0 K, only EAM3 and MEAM reproduce it correctly,
although, as shown in Section \ref{subsec:core}, the EAM3 potential does not
preserve this structure at high temperature.
In terms of $\sigma_P$, the five potentials studied here give a range of values
from 1.1 to 4.0 GPa. As Table \ref{tab:pot} shows, EAM2 and EAM3 display values
of 4.0 and 1.8 GPa, respectively. These potentials are also the most computationally 
efficient, which is an advantage when computing resources are limited. It is worth
noting that the cost of these potentials was evaluated using the cutoff radii
specified in the original references shown in Table \ref{tab:pot}.

Furthermore, the subspace of the energy landscape most relevant to
screw dislocation motion is the Peierls potential and the $\gamma$
surface (Figs.\ \ref{peierls} and \ref{fig:gamma}). The shape of the Peierls potential
is only correctly reproduced by potentials EAM3 and MEAM, while the rest
predict trajectories with metastable states along their path, in
disagreement with DFT calculations. Regarding the $\gamma$ surface,
potentials EAM1, EAM3 and MEAM all predict the essential qualitative and
quantitative features of the DFT results and are also in good agreement
with the results by Gr\"oger \etal~using a TB-BOP \cite{groger}.

Thus, on the basis of all these calculations, the MEAM potential appears to be the best
suited of those tried here to carry out dislocation simulations at any temperature. 
When computational cost is of the essence, EAM3 may be considered an acceptable replacement 
for static calculations or at low temperatures and stresses.

\subsection{Mobility of screw dislocations}
Dislocation mobility is highly multidimensional in that it displays
multiparametric dependencies, e.g.\ on stress, temperature, dislocation character, slip system, 
etc.  Dislocation velocities are difficult to infer from straining experiments,
while they are costly and subjected to size limitations in simulations.
Measurements \cite{schadler1964} and calculations \cite{gumbsch1999,li2002} of
edge dislocation velocities have been carried out in W. However, other than the values
computed by Tian and Woo \cite{tian2004} at very high stress ($>$3.6 GPa),
to our knowledge no data exist on screw dislocation mobility in W at low stresses.
In this work, we have focused on the temperature and stress dependences, while we have kept the 
dislocation character and the slip system fixed.

A quick look at Fig.\ \ref{fig:allpot} reveals several interesting details. First,
the EAM2 and MEAM consistently give the highest and lowest velocities,
respectively, regardless of temperature. Since atomistic simulations
commonly overestimate screw dislocation velocities, particularly in the
low-stress range, this may be another reason in favor of using MEAM.
This is likely to be due to the fact that the dislocation core remains
compact in the entire temperature range using the MEAM potential
(cf.\ Fig.\ \ref{fig:core}).
Second, screw dislocations move by thermally-activated mechanisms below
the Peierls stress, transitioning to a viscous damping regime above it.
At the maximum applied stress of 2000 MPa, some dislocations
have been driven past the Peierls stress as given by their respective potentials
(cf.\ Table \ref{tab:pot}).
This is certainly the case for the TF-BOP and possibly potentials EAM2 and EAM3.
One would therefore expect to see a gradual exhaustion of the thermally-activated
regime and a transition into a linear regime. Interestingly, such a transition
appears to occur for potential EAM1 which has $\sigma_P=4.0$ GPa. It was
shown by Gilbert \etal~\cite{gilbert2011}, however,
that the actual transition stress decreases with
the square of the temperature, which may be what is seen here.
Appropriate exponential fits to the data shown in Fig.\ \ref{fig:allpot}
carried out in the Appendix reveal useful parameters that define the thermally activated regime.

\subsection{Dislocation core transformation}\label{subsec:4.2}
The behavior that emanates from the results in Figs.\ \ref{fig:allpot} and \ref{fig:random}
for potential EAM3 is the manifestation of a temperature-driven dislocation core transformation
that occurs as a result of changes to the free energy landscape.
We have characterized this transformation via differential displacement maps
of time-averaged atomic positions at finite temperatures, and a quasiharmonic analysis of the
location of inflection points in the $\langle111\rangle$ interaction energy, which
is known to control the dislocation core structure
(cf.\ Figs.\ \ref{fig:random} and \ref{core_atability_withT}).

This is seen to affect the dislocation mobility as well. 
We have shown that the reported core transformation
has an impact on both the stress and temperature dependencies. Indeed, in the analysis
carried out in the Appendix, it is shown that the temperature
dependence of fitted $\sigma$-$v$ relations at $T<1200$ K for EAM3 does not carry over
to higher temperatures.
By contrast, the same analysis does not yield significant differences
between the low and high temperature regimes for potentials EAM1 and MEAM.
This is further indication that the core structure may impact the
motion mechanisms in the corresponding temperature range.

We emphasize, however, that as long as there does not exist independent evidence of this
dislocation core structural change with temperature, the discussion about its true impact on the 
dynamic behavior of screw dislocations remains solely speculative, and we cautiously warn against 
using the EAM3 potential above the observed transformation temperature of $\approx$1500 K.
In this sense, the quasi-harmonic analysis performed in Section
\ref{mark} would be very amenable to DFT calculations, as it consists solely of zero-temperature calculations. This would provide an
independent means to prove or disprove --at least within the limitations of the quasi-harmonic analysis-- 
the behavior predicted by EAM3 at high temperature.

\subsection{Mechanism of motion}
The dislocation trajectories shown in Fig.\ \ref{fig:traj} demonstrate that screw dislocations move primarily along the direction of the applied stress. The only notable exception is the TF-BOP, for which significant transitions out of plane are observed. The figure, however, does not provide insights into the atomistic mechanism of motion. Then, in Fig.\ \ref{fig:random} trajectories for the EAM3 were analyzed with higher spatial and temporal resolution at temperatures 600 and 1800 K. At both temperatures, the dislocation moves by elementary $\{110\}$ kink-pair episodes. It is reasonable to assume that this mechanism can be extrapolated to other potentials that yield similar effective trajectories (close to MRSS plane). However, for the EAM3 results, there are some differences in terms of the temperature at which the trajectory was extracted. At 1800 K it appears as though the unit mechanism is composed of one $+30^{\circ}$ jump (on a $(01\bar{1})$ plane) followed by a correlated $-30^{\circ}$ jump (on a $(10\bar{1})$ plane). In other words, the dislocation appears to move by kink-pair episodes on the $(11\bar{2})$ plane that consist of two alternating and correlated $\pm30^{\circ}$ kinks. This is consistent with the mechanism proposed by Duesbury \cite{duesbury}. Overall, this results in a trajectory that follows a random walk and that, on average, forms zero degrees with the MRSS plane. Interestingly, at 600 K kink pairs on the $+30^{\circ}$ plane seem to be favored in a proportion of three or more to one over $-30^{\circ}$ ones.  It is unclear at this point if the dislocation core transition discussed above for EAM3 is responsible for this difference. Again, as stated in Section \ref{subsec:4.2}, we are reluctant to construe this as real physical behavior until more is known about the core structure transformation. Our main message from the analysis of trajectories is that despite the MRSS plane being of the $\{112\}$ family, motion proceeds by way of kink pairs on $\{110\}$ planes presumably for all potentials.

\section{Summary}
To summarize, the main findings of this paper can be condensed into the
following main items:
\begin{itemize}
\item We have calculated static properties relevant to screw dislocations
  using five different interatomic potentials for W. These include three EAM,
  one BOP and one MEAM.
\item We have calculated screw dislocation mobilities for all potentials on a $\{112\}$
  glide plane. Our calculations provide elements to judge the MEAM potential
  as the most suitable for dislocation calculations.
\item We have observed a temperature-induced dislocation core transformation --from compact to dissociated-- for one of the potentials tested. Lacking independent confirmation, we cannot confirm whether this corresponds to a real physical phenomenon or is an artifact, but the transformation is indeed seen to impact the dynamic properties of dislocations.
\item Our analysis of the five interatomic potentials suggests, first, that the atomistic nature of the dislocation core governs behavior at larger scales
  and, second, a purely static treatment of the dislocation core is insufficient to predict and describe the dynamics of dislocations.
\end{itemize}

\section*{Acknowledgments}
This work was performed under the auspices of the U.S. Department of Energy by
Lawrence Livermore National Laboratory under Contract DE-AC52-07NA27344.
We specifically acknowledge support from the Laboratory
Directed Research and Development Program under project 11-ERD-023.
D.C. and J.M.P. acknowledge support from the 7$^{\mathrm{th}}$ Framework
Programme with project HiPER: European High Power Laser Energy Research
Facility, Grant Agreement No.~211737. We specifically acknowledge the PhD
program support from Universidad Polit\'ecnica de Madrid. This work was also partly funded by the RCUK Energy Programme under grant EP/I501045 and the European Communities under the contract of Association between EURATOM and CCFE. The views and opinions expressed herein do not necessarily reflect those of the European Commission.

\appendix
\section{\label{app}}

Here we analyze the overall impact of the dislocation core transition
for potential EAM3 on dislocation mobility.
We fit the data given in Fig.\ \ref{fig:moball} to the general expression:
\begin{equation}
v(\sigma,T)=A\sigma\exp{\left(-\frac{H_0-\sigma V^{\ast}}{kT}\right)}
\label{eq:app}
\end{equation}
where $A$, $H_0$, and $V^{\ast}$ are fitting constants that represent, respectively, a velocity
prefactor, the kink pair energy at 0 K, and the activation volume. We obtain these for
three potentials,
namely, EAM1, EAM3, and MEAM, and carry out the fit first including all temperatures. The
results for each case are shown in Table \ref{tab:app}.
\begin{table}[ht]
\caption{Fitting parameters for the analytical mobility function \ref{eq:app}. The average
fitting error for $A$, $H_0$, and $V^{\ast}$ was, respectively, 6, 9 and 10\%. Regular script:
values from full temperature fits; bold script: values from low temperature ($\leq900$ K) fits;
in parentheses: percentage difference between both sets of fits.}  
\centering
\begin{tabular}{c|c|c|c} 
\hline\hline 
Potential \T & $A$ $\left[{\mathrm{ms}}^{-1}\mathrm{MPa}^{-1}\right]$ & $H_0$ [eV] & $V^{\ast}$ $\left[b^3\right]$ \\ [0.5ex] 
\hline 
EAM3 \T & 0.26  {\bf 0.24} (8\%)  & 0.05  {\bf 0.04} (20\%) & 0.42  {\bf 0.19} (55\%) \\ 
MEAM & 0.19  {\bf 0.17} (10\%) & 0.08  {\bf 0.07} (12\%) & 0.26  {\bf 0.23} (12\%) \\
EAM1 & 0.30  {\bf 0.28} (7\%)  & 0.08  {\bf 0.07} (12\%) & 0.24  {\bf 0.24} (0\%) \\
\hline 
\end{tabular}
\label{tab:app} 
\end{table}
These values deserve some commentary, particularly $H_0$ and $V^{\ast}$.
Using the method described by Ventelon and Willaime \cite{ventelon2007}, we have obtained
a kink-pair energy of $H_0=$1.7 eV for the MEAM potential.
Experimentally, Brunner \cite{brunner2000} has obtained a value of 1.75 eV
from the temperature dependence of flow stress measurements in W, in very good agreement
with the calculated value but significantly higher than the MD values.
Giannattasio \etal~\cite{giannattasio2007}
have obtained values of the order of 1.0 eV inferred from the strain rate
dependence of the brittle-to-ductile transition, still much larger than those reported here.
Similarly, Tarleton and Roberts \cite{tarleton} have found values of $V^{\ast}$=20$b^3$ to be
representative of the kink-pair process in W.
Again, these are two orders of magnitude larger than ours.
The magnitudes of $H_0$ and $V^{\ast}$ obtained in our analysis suggest a
very `soft' thermally activated process, something not necessarily
consistent with the static data presented in Section \ref{subsec:pot}. The
low values of $H_0$ and  $V^{\ast}$ obtained in our simulations are
likely due to overdriven screw dislocation dynamics in the MD simulations.

Next, we obtain additional fits using only data at 300, 600, and 900 K,
{\it i.e.} at temperatures below the presumed core transformation for potentail EAM3.
The corresponding parameter values are shown in bold script for each case in Table
\ref{tab:app}. The percentage difference between the values for full and low temperature
fits is given in parentheses. Although the differences in the parameter $A$ are similar in all
cases, those for $H_0$ and $V^{\ast}$ are clearly largest for EAM3. Particularly, the differences
in the activation volume, which is known to be most sensitive to the core structure, are
considerable. This further reinforces
the notion that, from a dynamical standpoint, dislocations are behaving differently
below and above $\approx$1000 K. In contrast, for the other two potentials, the dynamic behavior
above and below this presumed transition temperature is governed by the same laws.
The low temperature fits obtained here are shown in Fig.\ \ref{fig:allpot}.
The fits provide very good agreement with the EAM1 and MEAM data at all temperatures,
whereas they gradually worsen for EAM3 as temperature increases.
As well, the deviation of the fits above $\sigma_P$ for potential EAM3 can be clearly appreciated.



\end{document}